\documentclass[twocolumn,longauth]{aa}  
\usepackage{graphicx}
\usepackage{txfonts}
\usepackage{amsmath,amsfonts,amssymb}
\usepackage{fixltx2e}
\usepackage{caption,subcaption}
\usepackage[breaklinks, colorlinks, citecolor=blue]{hyperref}
\usepackage{color}
\usepackage{natbib}
\bibpunct{(}{)}{;}{a}{}{,}

\def\setsymbol#1#2{\expandafter\def\csname #1\endcsname{#2}}
\def\getsymbol#1{\csname #1\endcsname}

\def\Planck{\textit{Planck}}

\def\alltwentyfifteenresultspapers{\nocite{planck2014-a01, planck2014-a03, planck2014-a04, planck2014-a05, planck2014-a06, planck2014-a07, planck2014-a08, planck2014-a09, planck2014-a11, planck2014-a12, planck2014-a13, planck2014-a14, planck2014-a15, planck2014-a16, planck2014-a17, planck2014-a18, planck2014-a19, planck2014-a20, planck2014-a22, planck2014-a24, planck2014-a26, planck2014-a28, planck2014-a29, planck2014-a30, planck2014-a31, planck2014-a35, planck2014-a36, planck2014-a37, planck2015-ES}}

\newbox\tablebox    \newdimen\tablewidth
\def\leaderfil{\leaders\hbox to 5pt{\hss.\hss}\hfil}
\def\endPlancktable{\tablewidth=\columnwidth 
    $$\hss\copy\tablebox\hss$$
    \vskip-\lastskip\vskip -2pt}

\def\tablenote#1 #2\par{\begingroup \parindent=0.8em
    \abovedisplayshortskip=0pt\belowdisplayshortskip=0pt
    \noindent
    $$\hss\vbox{\hsize\tablewidth \hangindent=\parindent \hangafter=1 \noindent
    \hbox to \parindent{$^#1$\hss}\strut#2\strut\par}\hss$$
    \endgroup}
\def\doubleline{\vskip 3pt\hrule \vskip 1.5pt \hrule \vskip 5pt}

\def\L2{\ifmmode L_2\else $L_2$\fi}

\def\DeltaT{\ifmmode \Delta T\else $\Delta T$\fi}
\def\deltat{\ifmmode \Delta t\else $\Delta t$\fi}
\def\fknee{\ifmmode f_{\rm knee}\else $f_{\rm knee}$\fi}
\def\Fmax{\ifmmode F_{\rm max}\else $F_{\rm max}$\fi}
\def\solar{\ifmmode{\rm M}_{\mathord\odot}\else${\rm M}_{\mathord\odot}$\fi}
\def\Msolar{\ifmmode{\rm M}_{\mathord\odot}\else${\rm M}_{\mathord\odot}$\fi}
\def\Lsolar{\ifmmode{\rm L}_{\mathord\odot}\else${\rm L}_{\mathord\odot}$\fi}
\def\inv{\ifmmode^{-1}\else$^{-1}$\fi}
\def\mo{\ifmmode^{-1}\else$^{-1}$\fi}
\def\sup#1{\ifmmode ^{\rm #1}\else $^{\rm #1}$\fi}
\def\expo#1{\ifmmode \times 10^{#1}\else $\times 10^{#1}$\fi}
\def\,{\thinspace}
\def\lsim{\mathrel{\raise .4ex\hbox{\rlap{$<$}\lower 1.2ex\hbox{$\sim$}}}}
\def\gsim{\mathrel{\raise .4ex\hbox{\rlap{$>$}\lower 1.2ex\hbox{$\sim$}}}}

\def\simprop{\mathrel{\raise .4ex\hbox{\rlap{$\propto$}\lower 1.2ex\hbox{$\sim$}}}}
\def\deg{\ifmmode^\circ\else$^\circ$\fi}
\def\pdeg{\ifmmode $\setbox0=\hbox{$^{\circ}$}\rlap{\hskip.11\wd0 .}$^{\circ}
          \else \setbox0=\hbox{$^{\circ}$}\rlap{\hskip.11\wd0 .}$^{\circ}$\fi}
\def\arcs{\ifmmode {^{\scriptstyle\prime\prime}}
          \else $^{\scriptstyle\prime\prime}$\fi}
\def\arcm{\ifmmode {^{\scriptstyle\prime}}
          \else $^{\scriptstyle\prime}$\fi}
\newdimen\sa  \newdimen\sb
\def\parcs{\sa=.07em \sb=.03em
     \ifmmode \hbox{\rlap{.}}^{\scriptstyle\prime\kern -\sb\prime}\hbox{\kern -\sa}
     \else \rlap{.}$^{\scriptstyle\prime\kern -\sb\prime}$\kern -\sa\fi}
\def\parcm{\sa=.08em \sb=.03em
     \ifmmode \hbox{\rlap{.}\kern\sa}^{\scriptstyle\prime}\hbox{\kern-\sb}
     \else \rlap{.}\kern\sa$^{\scriptstyle\prime}$\kern-\sb\fi}
\def\ra[#1 #2 #3.#4]{#1\sup{h}#2\sup{m}#3\sup{s}\llap.#4}
\def\dec[#1 #2 #3.#4]{#1\deg#2\arcm#3\arcs\llap.#4}
\def\deco[#1 #2 #3]{#1\deg#2\arcm#3\arcs}
\def\rra[#1 #2]{#1\sup{h}#2\sup{m}}

\def\dots{\relax\ifmmode \ldots\else $\ldots$\fi}
\def\WHzsr{\ifmmode $W\,Hz\mo\,sr\mo$\else W\,Hz\mo\,sr\mo\fi}
\def\mHz{\ifmmode $\,mHz$\else \,mHz\fi}
\def\GHz{\ifmmode $\,GHz$\else \,GHz\fi}
\def\mKs{\ifmmode $\,mK\,s$^{1/2}\else \,mK\,s$^{1/2}$\fi}
\def\muKs{\ifmmode \,\mu$K\,s$^{1/2}\else \,$\mu$K\,s$^{1/2}$\fi}
\def\muKRJs{\ifmmode \,\mu$K$_{\rm RJ}$\,s$^{1/2}\else \,$\mu$K$_{\rm RJ}$\,s$^{1/2}$\fi}
\def\muKHz{\ifmmode \,\mu$K\,Hz$^{-1/2}\else \,$\mu$K\,Hz$^{-1/2}$\fi}
\def\MJysr{\ifmmode \,$MJy\,sr\mo$\else \,MJy\,sr\mo\fi}
\def\MJysrmK{\ifmmode \,$MJy\,sr\mo$\,mK$_{\rm CMB}\mo\else \,MJy\,sr\mo\,mK$_{\rm CMB}\mo$\fi}
\def\microns{\ifmmode \,\mu$m$\else \,$\mu$m\fi}

\def\muK{\ifmmode \,\mu$K$\else \,$\mu$\hbox{K}\fi}
\def\microK{\ifmmode \,\mu$K$\else \,$\mu$\hbox{K}\fi}
\def\muW{\ifmmode \,\mu$W$\else \,$\mu$\hbox{W}\fi}
\def\kms{\ifmmode $\,km\,s$^{-1}\else \,km\,s$^{-1}$\fi}
\def\kmsMpc{\ifmmode $\,\kms\,Mpc\mo$\else \,\kms\,Mpc\mo\fi}
\providecommand{\sorthelp}[1]{}

\newcommand{\Euclid}{{\em Euclid}}

\newcommand{\PSZtwo}{{\tt PSZ2}}

\newcommand{\Ho}{H_0}
\newcommand{\OmM}{\Omega_{\rm m}}
\newcommand{\As}{A_{\rm s}}
\newcommand{\rhoc}{\rho_{\rm c}}

\newcommand{\snr}{S/N}
\newcommand{\sigmaf}{\sigma_{\rm f}}
\newcommand{\Yfive}{Y_{500}}
\newcommand{\meanYfive}{\bar{Y}_{500}}

\newcommand{\siglnY}{\sigma_{\rm lnY}}
\newcommand{\Rfive}{R_{500}}
\newcommand{\Mfive}{M_{500}}
\newcommand{\thetafive}{\theta_{500}}
\newcommand{\meanthetafive}{\bar{\theta}_{500}}
\newcommand{\Dang}{D_{\rm A}}
\newcommand{\qcat}{q_{\rm cat}}
\newcommand{\qm}{q_{\rm m}}
\newcommand{\meanqm}{\bar{q}_{\rm m}}

\newcommand{\Mlens}{M_{\rm lens}}
\newcommand{\MYsz}{M_{\rm Yz}}
\newcommand{\Mx}{M_{\rm X}}

\begin{document} 

   \title{\Planck\ 2015 results. XXIV. Cosmology from Sunyaev-Zeldovich cluster counts}
   \titlerunning{Cosmology from SZ cluster counts}
   
\author{\small
Planck Collaboration: P.~A.~R.~Ade\inst{97}
\and
N.~Aghanim\inst{65}
\and
M.~Arnaud\inst{81}
\and
M.~Ashdown\inst{77, 6}
\and
J.~Aumont\inst{65}
\and
C.~Baccigalupi\inst{95}
\and
A.~J.~Banday\inst{108, 11}
\and
R.~B.~Barreiro\inst{72}
\and
J.~G.~Bartlett\inst{1, 74}
\and
N.~Bartolo\inst{33, 73}
% ...
}
\institute{\small
APC, AstroParticule et Cosmologie, Universit\'{e} Paris Diderot, CNRS/IN2P3, CEA/lrfu, Observatoire de Paris, Sorbonne Paris Cit\'{e}, 10, rue Alice Domon et L\'{e}onie Duquet, 75205 Paris Cedex 13, France\goodbreak
\and
Aalto University Mets\"{a}hovi Radio Observatory and Dept of Radio Science and Engineering, P.O. Box 13000, FI-00076 AALTO, Finland\goodbreak
\and
African Institute for Mathematical Sciences, 6-8 Melrose Road, Muizenberg, Cape Town, South Africa\goodbreak
\and
Agenzia Spaziale Italiana Science Data Center, Via del Politecnico snc, 00133, Roma, Italy\goodbreak
\and
Aix Marseille Universit\'{e}, CNRS, LAM (Laboratoire d'Astrophysique de Marseille) UMR 7326, 13388, Marseille, France\goodbreak
\and
Astrophysics Group, Cavendish Laboratory, University of Cambridge, J J Thomson Avenue, Cambridge CB3 0HE, U.K.\goodbreak
\and
Astrophysics \& Cosmology Research Unit, School of Mathematics, Statistics \& Computer Science, University of KwaZulu-Natal, Westville Campus, Private Bag X54001, Durban 4000, South Africa\goodbreak
\and
Atacama Large Millimeter/submillimeter Array, ALMA Santiago Central Offices, Alonso de Cordova 3107, Vitacura, Casilla 763 0355, Santiago, Chile\goodbreak
\and
CGEE, SCS Qd 9, Lote C, Torre C, 4$^{\circ}$ andar, Ed. Parque Cidade Corporate, CEP 70308-200, Bras\'{i}lia, DF, Brazil\goodbreak
\and
CITA, University of Toronto, 60 St. George St., Toronto, ON M5S 3H8, Canada\goodbreak
\and
CNRS, IRAP, 9 Av. colonel Roche, BP 44346, F-31028 Toulouse cedex 4, France\goodbreak
\and
CRANN, Trinity College, Dublin, Ireland\goodbreak
\and
California Institute of Technology, Pasadena, California, U.S.A.\goodbreak
\and
Centre for Theoretical Cosmology, DAMTP, University of Cambridge, Wilberforce Road, Cambridge CB3 0WA, U.K.\goodbreak
\and
Centro de Estudios de F\'{i}sica del Cosmos de Arag\'{o}n (CEFCA), Plaza San Juan, 1, planta 2, E-44001, Teruel, Spain\goodbreak
\and
Computational Cosmology Center, Lawrence Berkeley National Laboratory, Berkeley, California, U.S.A.\goodbreak
\and
Consejo Superior de Investigaciones Cient\'{\i}ficas (CSIC), Madrid, Spain\goodbreak
\and
DSM/Irfu/SPP, CEA-Saclay, F-91191 Gif-sur-Yvette Cedex, France\goodbreak
\and
DTU Space, National Space Institute, Technical University of Denmark, Elektrovej 327, DK-2800 Kgs. Lyngby, Denmark\goodbreak
\and
D\'{e}partement de Physique Th\'{e}orique, Universit\'{e} de Gen\`{e}ve, 24, Quai E. Ansermet,1211 Gen\`{e}ve 4, Switzerland\goodbreak
\and
Departamento de Astrof\'{i}sica, Universidad de La Laguna (ULL), E-38206 La Laguna, Tenerife, Spain\goodbreak
\and
Departamento de F\'{\i}sica, Universidad de Oviedo, Avda. Calvo Sotelo s/n, Oviedo, Spain\goodbreak
\and
Department of Astronomy and Astrophysics, University of Toronto, 50 Saint George Street, Toronto, Ontario, Canada\goodbreak
\and
Department of Astrophysics/IMAPP, Radboud University Nijmegen, P.O. Box 9010, 6500 GL Nijmegen, The Netherlands\goodbreak
\and
Department of Physics \& Astronomy, University of British Columbia, 6224 Agricultural Road, Vancouver, British Columbia, Canada\goodbreak
\and
Department of Physics and Astronomy, Dana and David Dornsife College of Letter, Arts and Sciences, University of Southern California, Los Angeles, CA 90089, U.S.A.\goodbreak
\and
Department of Physics and Astronomy, University College London, London WC1E 6BT, U.K.\goodbreak
\and
Department of Physics, Florida State University, Keen Physics Building, 77 Chieftan Way, Tallahassee, Florida, U.S.A.\goodbreak
\and
Department of Physics, Gustaf H\"{a}llstr\"{o}min katu 2a, University of Helsinki, Helsinki, Finland\goodbreak
\and
Department of Physics, Princeton University, Princeton, New Jersey, U.S.A.\goodbreak
\and
Department of Physics, University of California, Santa Barbara, California, U.S.A.\goodbreak
\and
Department of Physics, University of Illinois at Urbana-Champaign, 1110 West Green Street, Urbana, Illinois, U.S.A.\goodbreak
\and
Dipartimento di Fisica e Astronomia G. Galilei, Universit\`{a} degli Studi di Padova, via Marzolo 8, 35131 Padova, Italy\goodbreak
\and
Dipartimento di Fisica e Scienze della Terra, Universit\`{a} di Ferrara, Via Saragat 1, 44122 Ferrara, Italy\goodbreak
\and
Dipartimento di Fisica, Universit\`{a} La Sapienza, P. le A. Moro 2, Roma, Italy\goodbreak
\and
Dipartimento di Fisica, Universit\`{a} degli Studi di Milano, Via Celoria, 16, Milano, Italy\goodbreak
\and
Dipartimento di Fisica, Universit\`{a} degli Studi di Trieste, via A. Valerio 2, Trieste, Italy\goodbreak
\and
Dipartimento di Matematica, Universit\`{a} di Roma Tor Vergata, Via della Ricerca Scientifica, 1, Roma, Italy\goodbreak
\and
Discovery Center, Niels Bohr Institute, Blegdamsvej 17, Copenhagen, Denmark\goodbreak
\and
Discovery Center, Niels Bohr Institute, Copenhagen University, Blegdamsvej 17, Copenhagen, Denmark\goodbreak
\and
European Southern Observatory, ESO Vitacura, Alonso de Cordova 3107, Vitacura, Casilla 19001, Santiago, Chile\goodbreak
\and
European Space Agency, ESAC, Planck Science Office, Camino bajo del Castillo, s/n, Urbanizaci\'{o}n Villafranca del Castillo, Villanueva de la Ca\~{n}ada, Madrid, Spain\goodbreak
\and
European Space Agency, ESTEC, Keplerlaan 1, 2201 AZ Noordwijk, The Netherlands\goodbreak
\and
Gran Sasso Science Institute, INFN, viale F. Crispi 7, 67100 L'Aquila, Italy\goodbreak
\and
HGSFP and University of Heidelberg, Theoretical Physics Department, Philosophenweg 16, 69120, Heidelberg, Germany\goodbreak
\and
Haverford College Astronomy Department, 370 Lancaster Avenue, Haverford, Pennsylvania, U.S.A.\goodbreak
\and
Helsinki Institute of Physics, Gustaf H\"{a}llstr\"{o}min katu 2, University of Helsinki, Helsinki, Finland\goodbreak
\and
INAF - Osservatorio Astrofisico di Catania, Via S. Sofia 78, Catania, Italy\goodbreak
\and
INAF - Osservatorio Astronomico di Padova, Vicolo dell'Osservatorio 5, Padova, Italy\goodbreak
\and
INAF - Osservatorio Astronomico di Roma, via di Frascati 33, Monte Porzio Catone, Italy\goodbreak
\and
INAF - Osservatorio Astronomico di Trieste, Via G.B. Tiepolo 11, Trieste, Italy\goodbreak
\and
INAF/IASF Bologna, Via Gobetti 101, Bologna, Italy\goodbreak
\and
INAF/IASF Milano, Via E. Bassini 15, Milano, Italy\goodbreak
\and
INFN, Sezione di Bologna, Via Irnerio 46, I-40126, Bologna, Italy\goodbreak
\and
INFN, Sezione di Roma 1, Universit\`{a} di Roma Sapienza, Piazzale Aldo Moro 2, 00185, Roma, Italy\goodbreak
\and
INFN, Sezione di Roma 2, Universit\`{a} di Roma Tor Vergata, Via della Ricerca Scientifica, 1, Roma, Italy\goodbreak
\and
INFN/National Institute for Nuclear Physics, Via Valerio 2, I-34127 Trieste, Italy\goodbreak
\and
IPAG: Institut de Plan\'{e}tologie et d'Astrophysique de Grenoble, Universit\'{e} Grenoble Alpes, IPAG, F-38000 Grenoble, France, CNRS, IPAG, F-38000 Grenoble, France\goodbreak
\and
ISDC, Department of Astronomy, University of Geneva, ch. d'Ecogia 16, 1290 Versoix, Switzerland\goodbreak
\and
IUCAA, Post Bag 4, Ganeshkhind, Pune University Campus, Pune 411 007, India\goodbreak
\and
Imperial College London, Astrophysics group, Blackett Laboratory, Prince Consort Road, London, SW7 2AZ, U.K.\goodbreak
\and
Infrared Processing and Analysis Center, California Institute of Technology, Pasadena, CA 91125, U.S.A.\goodbreak
\and
Institut N\'{e}el, CNRS, Universit\'{e} Joseph Fourier Grenoble I, 25 rue des Martyrs, Grenoble, France\goodbreak
\and
Institut Universitaire de France, 103, bd Saint-Michel, 75005, Paris, France\goodbreak
\and
Institut d'Astrophysique Spatiale, CNRS, Univ. Paris-Sud, Universit\'{e} Paris-Saclay, B\^{a}t. 121, 91405 Orsay cedex, France\goodbreak
\and
Institut d'Astrophysique de Paris, CNRS (UMR7095), 98 bis Boulevard Arago, F-75014, Paris, France\goodbreak
\and
Institut f\"ur Theoretische Teilchenphysik und Kosmologie, RWTH Aachen University, D-52056 Aachen, Germany\goodbreak
\and
Institute for Space Sciences, Bucharest-Magurale, Romania\goodbreak
\and
Institute of Astronomy, University of Cambridge, Madingley Road, Cambridge CB3 0HA, U.K.\goodbreak
\and
Institute of Theoretical Astrophysics, University of Oslo, Blindern, Oslo, Norway\goodbreak
\and
Instituto de Astrof\'{\i}sica de Canarias, C/V\'{\i}a L\'{a}ctea s/n, La Laguna, Tenerife, Spain\goodbreak
\and
Instituto de F\'{\i}sica de Cantabria (CSIC-Universidad de Cantabria), Avda. de los Castros s/n, Santander, Spain\goodbreak
\and
Istituto Nazionale di Fisica Nucleare, Sezione di Padova, via Marzolo 8, I-35131 Padova, Italy\goodbreak
\and
Jet Propulsion Laboratory, California Institute of Technology, 4800 Oak Grove Drive, Pasadena, California, U.S.A.\goodbreak
\and
Jodrell Bank Centre for Astrophysics, Alan Turing Building, School of Physics and Astronomy, The University of Manchester, Oxford Road, Manchester, M13 9PL, U.K.\goodbreak
\and
Kavli Institute for Cosmological Physics, University of Chicago, Chicago, IL 60637, USA\goodbreak
\and
Kavli Institute for Cosmology Cambridge, Madingley Road, Cambridge, CB3 0HA, U.K.\goodbreak
\and
Kazan Federal University, 18 Kremlyovskaya St., Kazan, 420008, Russia\goodbreak
\and
LAL, Universit\'{e} Paris-Sud, CNRS/IN2P3, Orsay, France\goodbreak
\and
LERMA, CNRS, Observatoire de Paris, 61 Avenue de l'Observatoire, Paris, France\goodbreak
\and
Laboratoire AIM, IRFU/Service d'Astrophysique - CEA/DSM - CNRS - Universit\'{e} Paris Diderot, B\^{a}t. 709, CEA-Saclay, F-91191 Gif-sur-Yvette Cedex, France\goodbreak
\and
Laboratoire Traitement et Communication de l'Information, CNRS (UMR 5141) and T\'{e}l\'{e}com ParisTech, 46 rue Barrault F-75634 Paris Cedex 13, France\goodbreak
\and
Laboratoire de Physique Subatomique et Cosmologie, Universit\'{e} Grenoble-Alpes, CNRS/IN2P3, 53, rue des Martyrs, 38026 Grenoble Cedex, France\goodbreak
\and
Laboratoire de Physique Th\'{e}orique, Universit\'{e} Paris-Sud 11 \& CNRS, B\^{a}timent 210, 91405 Orsay, France\goodbreak
\and
Lawrence Berkeley National Laboratory, Berkeley, California, U.S.A.\goodbreak
\and
Lebedev Physical Institute of the Russian Academy of Sciences, Astro Space Centre, 84/32 Profsoyuznaya st., Moscow, GSP-7, 117997, Russia\goodbreak
\and
Max-Planck-Institut f\"{u}r Astrophysik, Karl-Schwarzschild-Str. 1, 85741 Garching, Germany\goodbreak
\and
McGill Physics, Ernest Rutherford Physics Building, McGill University, 3600 rue University, Montr\'{e}al, QC, H3A 2T8, Canada\goodbreak
\and
National University of Ireland, Department of Experimental Physics, Maynooth, Co. Kildare, Ireland\goodbreak
\and
Nicolaus Copernicus Astronomical Center, Bartycka 18, 00-716 Warsaw, Poland\goodbreak
\and
Niels Bohr Institute, Blegdamsvej 17, Copenhagen, Denmark\goodbreak
\and
Niels Bohr Institute, Copenhagen University, Blegdamsvej 17, Copenhagen, Denmark\goodbreak
\and
Nordita (Nordic Institute for Theoretical Physics), Roslagstullsbacken 23, SE-106 91 Stockholm, Sweden\goodbreak
\and
Optical Science Laboratory, University College London, Gower Street, London, U.K.\goodbreak
\and
SISSA, Astrophysics Sector, via Bonomea 265, 34136, Trieste, Italy\goodbreak
\and
SMARTEST Research Centre, Universit\`{a} degli Studi e-Campus, Via Isimbardi 10, Novedrate (CO), 22060, Italy\goodbreak
\and
School of Physics and Astronomy, Cardiff University, Queens Buildings, The Parade, Cardiff, CF24 3AA, U.K.\goodbreak
\and
School of Physics and Astronomy, University of Nottingham, Nottingham NG7 2RD, U.K.\goodbreak
\and
Sorbonne Universit\'{e}-UPMC, UMR7095, Institut d'Astrophysique de Paris, 98 bis Boulevard Arago, F-75014, Paris, France\goodbreak
\and
Space Research Institute (IKI), Russian Academy of Sciences, Profsoyuznaya Str, 84/32, Moscow, 117997, Russia\goodbreak
\and
Space Sciences Laboratory, University of California, Berkeley, California, U.S.A.\goodbreak
\and
Special Astrophysical Observatory, Russian Academy of Sciences, Nizhnij Arkhyz, Zelenchukskiy region, Karachai-Cherkessian Republic, 369167, Russia\goodbreak
\and
Stanford University, Dept of Physics, Varian Physics Bldg, 382 Via Pueblo Mall, Stanford, California, U.S.A.\goodbreak
\and
Sub-Department of Astrophysics, University of Oxford, Keble Road, Oxford OX1 3RH, U.K.\goodbreak
\and
The Oskar Klein Centre for Cosmoparticle Physics, Department of Physics,Stockholm University, AlbaNova, SE-106 91 Stockholm, Sweden\goodbreak
\and
Theory Division, PH-TH, CERN, CH-1211, Geneva 23, Switzerland\goodbreak
\and
UPMC Univ Paris 06, UMR7095, 98 bis Boulevard Arago, F-75014, Paris, France\goodbreak
\and
Universit\'{e} de Toulouse, UPS-OMP, IRAP, F-31028 Toulouse cedex 4, France\goodbreak
% ...
}

   \authorrunning{Planck Collaboration}
   
   \date{Received ; accepted}

 
\abstract{
We present cluster counts and corresponding cosmological constraints from the \Planck\ full mission data set.  Our catalogue consists of 439 clusters detected via their Sunyaev-Zeldovich (SZ) signal down to a signal-to-noise ratio of 6, and is more than a factor of 2 larger than the 2013 \Planck\ cluster cosmology sample.  The counts are consistent with those from 2013 and yield compatible  constraints under the same modelling assumptions. Taking advantage of the larger catalogue, we extend our analysis to the two-dimensional distribution in redshift and signal-to-noise.  We use mass estimates from two recent studies of gravitational lensing of background galaxies by \Planck\ clusters to provide priors on the hydrostatic bias parameter, $(1-b)$. In addition, we use lensing of cosmic microwave background (CMB) temperature fluctuations by \Planck\ clusters as an independent constraint on this parameter.  These various calibrations imply constraints on the present-day amplitude of matter fluctuations in varying degrees of tension with those from the \Planck\ analysis of primary fluctuations in the CMB; for the lowest estimated values of $(1-b)$ the tension is mild, only a little over one standard deviation, while it remains substantial ($3.7\,\sigma$) for the largest estimated value. We also examine constraints on extensions to the base flat $\Lambda$CDM model by combining the cluster and CMB constraints.  The combination appears to favour non-minimal neutrino masses, but this possibility does little to relieve the overall tension because it simultaneously lowers the implied value of the Hubble parameter, thereby exacerbating the discrepancy with most current astrophysical estimates. Improving the precision of cluster mass calibrations from the current 10\,\%-level to 1\,\% would significantly strengthen these combined analyses and provide a stringent test of the base $\Lambda$CDM model.
}




\maketitle

\alltwentyfifteenresultspapers

%

\clearpage

\section{Introduction}

Galaxy cluster counts are a standard cosmological tool that has found powerful application in recent Sunyaev-Zeldovich (SZ) surveys performed by the Atacama Cosmology Telescope \citep[ACT,][]{swetz2011,hasselfield2013}, the South Pole Telescope \citep[SPT,][]{carlstrom2011,benson2013,reichardt2013,bocquet2014}, and the \Planck\ satellite\footnote{\Planck\ (\url{http://www.esa.int/Planck}) is a project of the European Space Agency  (ESA) with instruments provided by two scientific consortia funded by ESA member states and led by Principal Investigators from France and Italy, telescope reflectors provided through a collaboration between ESA and a scientific consortium led and funded by Denmark, and additional contributions from NASA (USA).} \citep{tauber2010a,planck2011-1.1}.  The abundance of clusters and its evolution are sensitive to the cosmic matter density, $\OmM$, and the present amplitude of density fluctuations, characterized by $\sigma_8$,  the rms linear overdensity in spheres of radius $8h^{-1}$\,Mpc. The primary cosmic microwave background (CMB) anisotropies, on the other hand, reflect the density perturbation power spectrum at the time of recombination.  This difference is important because a comparison of the amplitude of the perturbations at the two epochs tests the evolution of density perturbations from recombination until today, enabling us to look for possible extensions to the base $\Lambda$CDM model, such as non-minimal neutrino masses or non-zero curvature.  

Launched on 14 May 2009, \Planck\ scanned the entire sky twice a year from 12~August 2009 to 23~October 2013, at angular resolutions from 33\arcm\ to 5\arcm with two instruments: the Low Frequency Instrument \citep[LFI;][]{Bersanelli2010, planck2011-1.4}, covering bands centred at 30, 44, and 70\,GHz, and the High Frequency Instrument \citep[HFI;][]{Lamarre2010, planck2011-1.5}, covering bands centred at 100, 143, 217, 353, 545, and $857\,$GHz.  

An initial set of cosmology results appeared in 2013, based on the first 15.5 months of data \citep{planck2013-p01}, including cosmological constraints from the redshift distribution of 189 galaxy clusters detected at  signal-to-noise (\snr)\ $> 7$~\citep[hereafter, our ``first analysis'' or the ``2013 analysis'',][]{planck2013-p15}.  The present paper is part of the second set of cosmology results obtained from the full mission data set; it is based on an updated cluster sample introduced in an accompanying paper \citep[the \PSZtwo,][]{planck2014-a01}.  

Our first analysis found fewer clusters than predicted by \Planck's base $\Lambda$CDM model, expressed as a tension between the cluster constraints on $(\OmM, \sigma_8)$ and those from the primary CMB anisotropies \citep{planck2013-p11}.  This could reflect the need for an extension to the base $\Lambda$CDM model or indicate that clusters are more massive than determined by the SZ signal-mass scaling relation adopted in 2013. 

The cluster mass scale is the largest source of uncertainty in interpretation of the cluster counts.  We based our first analysis on X-ray mass proxies that rely on the assumption of hydrostatic equilibrium.  Simulations demonstrate that this assumption can be violated by bulk motions in the gas or by non-thermal sources of pressure \citep[e.g., magnetic fields or cosmic rays,][]{2007ApJ...668....1N,2008A&A...491...71P,2010A&A...514A..93M}. Systematics in the X-ray analyses (e.g., instrument calibration, temperature structure in the gas) could also bias the mass measurements significantly \citep{rasia2006,rasia2012}.  We quantified our ignorance of the true mass scale of clusters with a mass bias parameter that was varied over the range $0-30\,\%$, with a baseline value of 20\,\% (see below for the definition of the mass bias), as suggested by numerical simulations \citep[see the Appendix of][]{planck2013-p15}.

Gravitational lensing studies of the SZ signal-mass relation are particularly valuable in this context because they are independent of the dynamical state of the cluster \citep{marrone2012,planck2012-III}, although they also, of course, can be affected by systematic effects \citep[e.g., ][]{becker2011}.  New, more precise lensing mass measurements for \Planck\ clusters have appeared since our 2013 analysis \citep[][]{vonderlinden2014, hoekstra2015}.  We incorporate these new results as prior constraints on the mass bias in the present analysis.  Two other improvements over 2013 are the use of a larger cluster catalogue and analysis of the counts in signal-to-noise as well as redshift.   

In addition, we apply a novel method to measure cluster masses through lensing of the CMB anisotropies.  This method, presented in \cite{melin2014}, enables us to use \Planck\ data alone to constrain the cluster mass scale.  It provides an important independent mass determination, which we compare to the galaxy lensing results, and one that is representative in the sense that it averages over the entire cluster cosmology sample, rather than a particularly chosen subsample. It is, however, less well tested than the other lensing methods because of its novel nature, and we comment on various potential systematics deserving further examination.

Our conventions throughout the paper are as follows.  We specify cluster mass, $\Mfive$, as the total mass within a sphere of radius $\Rfive$, defined as the radius within which the mean mass over-density of the cluster is 500 times the cosmic critical density at its redshift, $z$: $\Mfive = (4\pi/3) \Rfive^3 [500\rhoc(z)]$, with $\rhoc(z) = 3H^2(z)/(8\pi G)$, where $H(z)$ is the Hubble parameter with present-day value $\Ho = h\times 100\kmsMpc$.  We give SZ signal strength, $\Yfive$, in terms of the Compton $y$-profile integrated within a sphere of radius $\Rfive$, and we assume that all clusters follow the universal pressure profile of \citet{arnaud2010}.  Density parameters are defined relative to the present-day critical density, e.g., $\OmM = \rho_{\rm m}/\rhoc(z=0)$ for the total matter density, $\rho_{\rm m}$.  

We begin in the next section with a presentation of the \Planck\ 2015 cluster cosmology samples.  In Sect.~\ref{sec:modeling} we develop our model for the cluster counts in both redshift and signal-to-noise, including a discussion of the scaling relation, the scatter and the sample selection function.  Section~\ref{sec:cms} examines the overall cluster mass scale in light of recent gravitational lensing measurements; we also present our own calibration of the cluster mass scale based on lensing of the CMB temperature fluctuations.  Construction of the cluster likelihood and selection of external data sets is detailed in Sect.~\ref{sec:like}. 
We compare results based on our new likelihood to the 2013 \Planck\ cluster results in Sect.~\ref{sec:comparison2013}. We then present our 2015 cosmological constraints in Sect.~\ref{sec:constraints}, summarizing and discussing the results in Sect.~\ref{sec:discussion}.  We examine the potential impact of different modelling uncertainties in the Appendix.


\begin{figure}
\centering
\includegraphics[width=\hsize]{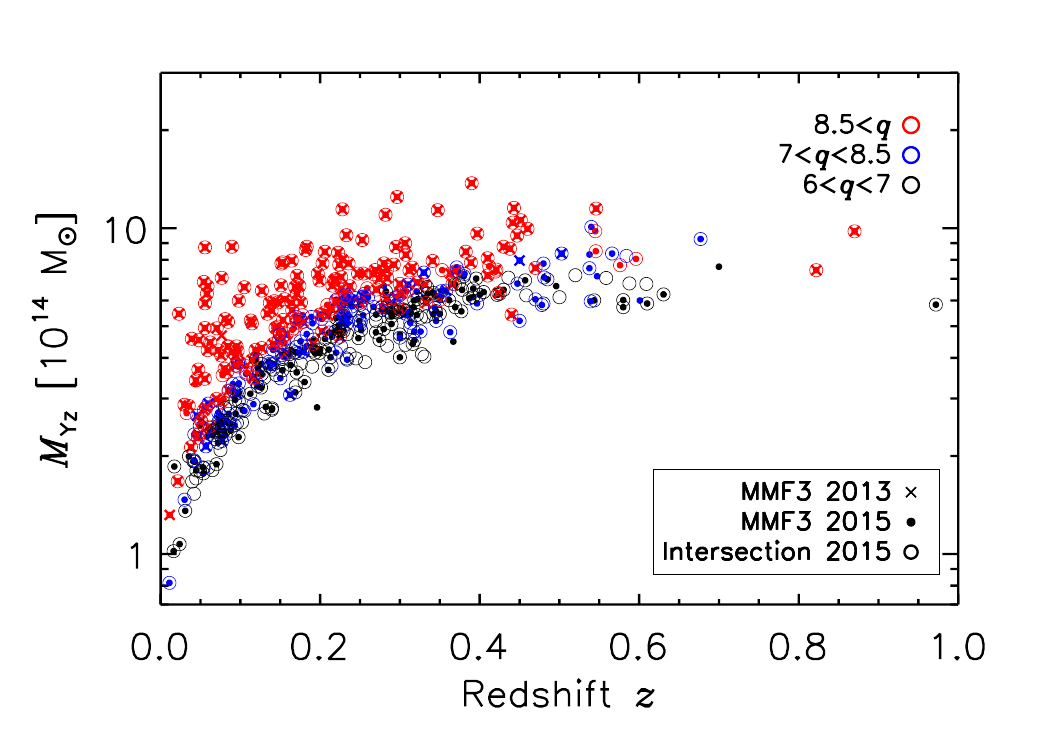}
\caption{Mass-redshift distribution of the \Planck\  cosmological samples colour-coded by their signal-to-noise, $q$.  The baseline MMF3 2015 cosmological sample is shown as the small filled circles. Objects which were in the MMF3 2013 cosmological sample are marked by crosses, while those in the 2015 intersection sample are shown as open circles.  The final samples are defined by $q>6$.  The mass $\MYsz$ is the \Planck\ mass proxy \citep[see text,][]{arnaud2015}. 
}
\label{fig:cosmo_samples}
\end{figure}

\section{The \Planck\ cosmological samples}
\label{sec:sampledef}

We detect clusters across the six highest frequency \Planck\ bands \citep[$100-857\GHz$,][]{planck2014-a08,planck2014-a09} using two implementations of the multi-frequency matched filter \citep[MMF3 and MMF1,][]{melin2006,planck2013-p05a} and a Bayesian extension \citep[PwS,][]{carvalho2009} that all incorporate the known (non-relativistic) SZ spectral signature and a model for the spatial profile of the signal.  The latter is taken to be the so-called ``universal pressure profile'' from \citet{arnaud2010} --- with the non-standard self-similar scaling --- and parameterized by an angular scale, $\thetafive$. 

We empirically characterize noise (all non-SZ signals) in localized sky patches ($10 \deg$ on a side for MMF3) using the set of cross-frequency power-spectra. We construct the filters with the resulting noise weights, and we then filter the set of six frequency maps over a range of cluster scales, $\thetafive$, spanning 1--35\,arcmin.  The filter returns an estimate of $\Yfive$ for each scale, based on the adopted profile template, and sources are finally assigned the $\thetafive$ (and hence $\Yfive$) value of the scale that maximizes their signal-to-noise.  Details are given in \cite{planck2013-p05a} and in an accompanying paper introducing the \Planck\ full-mission SZ catalogue \citep[\PSZtwo,][]{planck2014-a36}.

We define two cosmological samples from the general \PSZtwo\ catalogues, one consisting of detections by the MMF3 matched filter and the other of objects detected by all three methods (the intersection catalogue).  Both are defined by a signal-to-noise (denoted $q$ throughout) cut of $q  > 6$.  We then apply a mask to remove regions of high dust emission and point sources, leaving 65\,\% of the sky unmasked. The general catalogues, noise maps and masks can be downloaded from the Planck Legacy Archive.\footnote{\url{http://pla.esac.esa.int/pla/}}

The cosmological samples can be easily constructed from the \PSZtwo\ union and MMF3 catalogues.  The MMF3 cosmology sample is the subsample of the MMF3 catalogue defined by  $q > 6$ and for which the entry in the union catalogue has {\verb COSMO }='T'.
The intersection cosmology sample is defined from the union catalogue by the criteria {\verb COSMO }='T', {\verb PIPEDET }=111, and $q> 6$. 

Fig.~\ref{fig:cosmo_samples} shows the distribution of these samples in mass and redshift, together with the 2013 cosmology sample. The mass here is the \Planck\  mass proxy, $\MYsz$, defined in \citet{arnaud2015}~\citep[see also Section~7.2.2. in][]{planck2013-p05a} and taken from the \PSZtwo\ catalogue. It is calculated using the \Planck\ size-flux posterior contours in conjunction with X-ray priors to break the size-flux degeneracy inherent to the large \Planck\ beams \citep[see, e.g., Fig. 16 of][]{planck2014-a36}. The samples span masses in the range $(2-10)\times 10^{14} \, M_\odot$ and redshifts from $z=0$ to 1.\footnote{We fix $h=0.7$ and $\Omega_\Lambda=1-\OmM=0.7$ for the mass calculation.} This quantity$\MYsz$ is used in the external lensing mass calibration measurements, as discussed in Sect.~\ref{sec:cms}. 

The MMF3 (intersection) sample contains 439 (493) detections.  We note that the intersection catalogue has more objects than the MMF3 catalogue because of the different definitions of the signal-to-noise in the various catalogues. The signal-to-noise for the intersection catalogue corresponds to the highest signal-to-noise of the three detection algorithms (MMF1, MMF3, or PwS), while for the MMF3 catalogue we use its corresponding signal-to-noise.  As a consequence, the lowest value for the MMF3 signal-to-noise in the intersection sample is 4.8.  We note that, while being above our detection limit, the Virgo and the Perseus clusters are not part of our samples. This is because Virgo is too extended to be blindly detected by our algorithms and Perseus is close to a masked region.

The 2015 MMF3 cosmology sample contains all but one of the 189 clusters of the 2013 MMF3 sample. The missing cluster is PSZ1 980, which falls inside the 2015 point source mask. Six (14) redshifts are missing from the MMF3 (intersection) sample. Our analysis accounts for these by renormalizing the observed counts to redistribute the missing fraction uniformly across the considered redshift range~[0,1].  The small number of clusters with missing redshifts has no significant impact on our results.

We use the MMF3 cosmology sample at $q>6$ for our baseline analysis and the intersection sample for consistency checks, as detailed in the Appendix. In particular, we show that the intersection sample yields equivalent constraints.


\section{Modelling cluster counts}
\label{sec:modeling}
From the theoretical perspective, cluster abundance is a function of halo mass and redshift, as specified by the mass function.  Observationally, we detect clusters in \Planck\ through their SZ signal strength or, equivalently, their signal-to-noise and measure their redshift with follow-up observations.  The observed cluster counts are therefore a function of redshift, $z$, and signal-to-noise, $q$.   While we restricted our 2013 cosmology analysis to the redshift distribution alone \citep{planck2013-p15}, the larger catalogue afforded by the full mission data set offers the possibility of an analysis in both redshift and signal-to-noise.  We therefore develop the theory in terms of the joint distribution of clusters in the $(z,q)$-plane and then relate it to the more specific analysis of the redshift distribution to compare with our previous results.

\subsection{Counts as a function of redshift and signal-to-noise}
The distribution of clusters in redshift and and signal-to-noise can be written as
\begin{equation}
\label{eq:dndzdq}
\frac{dN}{dz dq} = \int d\Omega_{\rm mask} \int d\Mfive \, \frac{dN}{dz d\Mfive d\Omega}\, P[q | \meanqm(\Mfive,z,l,b)],
\end{equation}
with
\begin{equation}
\frac{dN}{dz d\Mfive d\Omega} = \frac{dN}{dV d\Mfive}\frac{dV}{dzd\Omega},
\end{equation}
i.e., the dark matter halo mass function times the volume element.  We adopt the mass function from \citet{tinker2008} throughout, apart from the Appendix, where we compare to the \citet{watson2013} mass function as a test of modelling robustness; there, we show that the \citet{watson2013} mass function yields constraints similar to those from the \citet{tinker2008} mass function, but shifted by about $1 \sigma$ towards higher $\OmM$ and lower $\sigma_8$ along the main degeneracy line.  

The quantity $P[q | \meanqm(\Mfive,z,l,b)]$ is the distribution of $q$ given the mean signal-to-noise value, $\meanqm(\Mfive,z,l,b)$, predicted by the model for a cluster of mass $\Mfive$ and redshift $z$ located at Galactic coordinates $(l,b)$.\footnote{This form assumes, as we do throughout, that the distribution depends on $z$ and $\Mfive$ only through the mean value $\meanqm$, specifically, that the intrinsic scatter, $\siglnY$, of Eq.~(\ref{eq:obsdist}) is constant.}  This latter quantity is defined as the ratio of the mean SZ signal expected of a cluster, $\meanYfive(\Mfive,z)$, as given in Eq.~(\ref{eq:Yscaling}), and the detection filter noise, $\sigmaf[\meanthetafive(\Mfive,z),l,b]$:
\begin{equation}
\label{eq:meanqm}
\meanqm \equiv \meanYfive(\Mfive,z)/\sigmaf[\meanthetafive(\Mfive,z),l,b].
\end{equation}
The filter noise depends on sky location ($l,b$) and the cluster angular size, $\meanthetafive$, which introduces additional dependence on mass and redshift. More detail on $\sigmaf$ can be found in~\cite{planck2013-p15} (see in particular figure~4 therein).

The distribution $P[q | \meanqm]$ incorporates noise fluctuations and intrinsic scatter in the actual cluster $\Yfive$ around the mean value, $\meanYfive(\Mfive,z)$, predicted from the scaling relation.  We discuss this scaling relation and our log-normal model for the intrinsic scatter below, and Sect.~\ref{sec:cms} examines the calibration of the overall mass scale for the scaling relation. 

The redshift distribution of clusters detected at $q > \qcat$ is the integral of Eq.~(\ref{eq:dndzdq}) over signal-to-noise,
\begin{eqnarray}
\label{eq:dndz}
\nonumber
\frac{dN}{dz}(q>\qcat) & = & \int_{\qcat}^\infty dq\, \frac{dN}{dz dq}  \\
                   & = & \int d\Omega \int d\Mfive \, \hat{\chi}(\Mfive,z,l,b) \, \frac{dN}{dz d\Mfive d\Omega}, \hspace{5mm}
\end{eqnarray}
with
\begin{equation}
\label{eq:chi14}
\hat{\chi} (\Mfive,z,l,b)  =  \int_{\qcat}^\infty dq\,  P[q | \meanqm(\Mfive,z,l,b)].
\end{equation}
Equation~(\ref{eq:dndz}) is equivalent to the expression used in our 2013 analysis if we write it in the form
\begin{equation}
\label{eq:chi13}
\hat{\chi} = \int d\ln\Yfive \int d\thetafive P(\ln\Yfive, \thetafive | z,\Mfive) \, \chi(\Yfive,\thetafive,l,b),  
\end{equation}
where $\chi(\Yfive,\thetafive,l,b)$ is the survey selection function at $q>\qcat$ in terms of true cluster parameters (Sect.~\ref{sec:sf}), and $P(\ln\Yfive, \thetafive | z,\Mfive)$ is the distribution of these parameters given cluster mass and redshift.  We specify the relation between Eq.~(\ref{eq:chi14}) and Eq.~(\ref{eq:chi13}) in the next section. 

\subsection{Observable-mass relations}
A crucial element of our modelling is the relation between cluster observables, $\Yfive$ and $\thetafive$, and halo mass and redshift.  Due to intrinsic variations in cluster properties, this relation is described by a distribution function, $P(\ln\Yfive,\thetafive | \Mfive, z)$, whose mean values are specified by the scaling relations $\meanYfive(\Mfive,z)$ and $\meanthetafive(\Mfive,z)$.  

We use the same form for these scaling relations as in our 2013 analysis:
\begin{equation}
\label{eq:Yscaling} 
E^{-\beta}(z)\left[\frac{\Dang^2(z) \meanYfive}{\mathrm{10^{-4}\,Mpc^2}}\right] =  Y_\ast \left[ {h \over 0.7}
  \right]^{-2+\alpha} \left[\frac{(1-b)\,
    \Mfive}{6\times10^{14}\,\Msolar}\right]^{\alpha},
\end{equation}
and
\begin{equation}
\label{eq:sizescaling}
{\meanthetafive}=\theta_\ast
\left[\frac{h}{0.7}\right]^{-2/3}\left[{(1-b)\,M_{500}\over 3\times
    10^{14} \Msolar}\right]^{1/3} \,E^{-2/3}(z)\,\left[\Dang(z)\over 500\,{\mathrm{Mpc}}\right]^{-1}\!\!,
\end{equation}
where $\theta_\ast=6.997\,$arcmin, and fiducial ranges for the parameters $Y_\ast$, $\alpha$, and $\beta$ are listed in Table~\ref{tab:params}; these values are identical to those used in our 2013 analysis.  Unless otherwise stated, we use Gaussian distributions with mean and standard deviation given by these values as prior constraints; one notable exception will be when we simultaneously fit for $\alpha$ and cosmological parameters.  In the above expressions, $\Dang(z)$ is the angular diameter distance and $E(z) \equiv H(z)/\Ho$.

\begin{table}[tb]
\begingroup
\newdimen\tblskip \tblskip=5pt
\caption{Summary of SZ-mass scaling-law parameters (see Eq.~\ref{eq:Yscaling}).}
  \label{tab:params}
\nointerlineskip
\vskip -3mm
\footnotesize
\setbox\tablebox=\vbox{
   \newdimen\digitwidth 
   \setbox0=\hbox{\rm 0} 
   \digitwidth=\wd0 
   \catcode`*=\active 
   \def*{\kern\digitwidth}
   \newdimen\signwidth 
   \setbox0=\hbox{+} 
   \signwidth=\wd0 
   \catcode`!=\active 
   \def!{\kern\signwidth}
\halign{#\hfil\tabskip=2em & \hfil#\tabskip=0pt\cr                           
\noalign{\doubleline}
                         Parameter & Value~~~~\cr
  \noalign{\vskip 3pt\hrule\vskip 5pt}
$\log Y_{\ast}$& $-0.19 \pm 0.02*$\cr
$\alpha^{\,\rm a}$ & $1.79\pm0.08*$\cr
$\beta^{\,\rm b}$ & $0.66\pm 0.50*$\cr
$\sigma_{\ln Y}^{\,\rm c}$ &$0.173 \pm 0.023$\cr                            
\noalign{\vskip 5pt\hrule\vskip 3pt}}}
\endPlancktable                    
\tablenote {{\rm a}} Except when specified, $\alpha$ is constrained by this prior in our one-dimensional likelihood over $N(z)$, but left free in our two-dimensional likelihood over $N(z,q)$.\par
\tablenote {{\rm b}} We fix $\beta$ to its central value throughout, except when examining modelling uncertainties in the Appendix.\par
\tablenote {{\rm c}} The value is the same as in our 2013 analysis, given here in terms of the natural logarithm and computed from $\sigma_{\log Y}=0.075 \pm 0.01$.\par
\endgroup
\end{table}                        

These scaling relations have been established by X-ray observations, as detailed in the Appendix of \citet{planck2013-p15}, and rely on mass determinations, $\Mx$, based on hydrostatic equilibrium of the intra-cluster gas.  The ``mass bias'' parameter, $b$, assumed to be constant in both mass and redshift, allows for any difference between the X-ray determined masses and true cluster halo mass: $\Mx=(1-b)\Mfive$. This is discussed at length in Sect.~\ref{sec:cms}. 

We adopt a log-normal\footnote{In this paper, ``$\ln$" denotes the natural logarithm and "$\log$" the logarithm to base 10; the expression is written in terms of the natural logarithm.} distribution for $\Yfive$ around its mean value $\meanYfive$, and a delta function for $\thetafive$ centred on $\meanthetafive$:
\begin{eqnarray}
\label{eq:obsdist}
\nonumber
P(\ln\Yfive, \thetafive | \Mfive, z)  & = & \frac{1}{\sqrt{2\pi}\siglnY}e^{-\ln^2(\Yfive/\meanYfive)/(2\siglnY^2)}\\
	                                      & \times & \delta[\thetafive-\meanthetafive],
\end{eqnarray}
where $\meanYfive(\Mfive,z)$ and $\meanthetafive(\Mfive,z)$ are given by Eqs.~(\ref{eq:Yscaling}) and (\ref{eq:sizescaling}). The $\delta$-function maintains the empirical definition of $\Rfive$ that is used in observational determinations of the profile. 

We can now specify the relation between Eqs.~(\ref{eq:chi14}) and (\ref{eq:chi13}) by noting that
\begin{equation}
\label{eq:qdist}
P[q | \meanqm(\Mfive,z,l,b)] = \int d\ln\qm P[q | \qm] P[\ln\qm | \meanqm],
\end{equation}
where $P[q | \qm]$ is the distribution of observed signal-to-noise, $q$, given the model value, $\qm$.  The second distribution represents intrinsic cluster scatter, which we write in terms of our observable-mass distribution, Eq.~({\ref{eq:obsdist}), as
\begin{eqnarray}
\label{eq:intscat}
\nonumber
P[\ln\qm | \meanqm] & = &\int d\thetafive P[\ln\Yfive(\ln\qm,\thetafive,l,b),\thetafive | \Mfive,z] \\
                               & = & \frac{1}{\sqrt{2\pi}\siglnY}e^{-\ln^2(\qm/\meanqm)/2\siglnY^2}.
\end{eqnarray}
Performing the integral of Eq.~(\ref{eq:chi14}), we find
\begin{equation}
\hat{\chi} = \int d\ln\qm P[\ln\qm | \meanqm] \chi(\Yfive,\thetafive,l,b),
\end{equation}
with the definition of our survey selection function 
\begin{equation}
\label{eq:sf}
\chi(\Yfive,\thetafive,l,b) = \int_{\qcat}^\infty dq P[q | \qm(\Yfive,\thetafive,l,b)].  
\end{equation}
We then reproduce Eq.~(\ref{eq:chi13})
by using the first line of Eq.~(\ref{eq:intscat}) and Eq.~(\ref{eq:meanqm}).

\subsection{Selection function and survey completeness}
\label{sec:sf}

The fundamental quantity describing the survey selection is $P[q |\qm]$, introduced in Eq.~(\ref{eq:qdist}).  It gives the observed signal-to-noise, used to select SZ sources, as a function of model (``true'') cluster parameters through $\qm(\Yfive, \thetafive,l,b)$, and it defines the ``survey selection function'' $\chi(\Yfive,\thetafive,l,b)$ via Eq.~(\ref{eq:sf}). We characterize the survey selection in two ways.  The first is with an analytical model and the second employs a Monte Carlo extraction of simulated sources injected into the \Planck\ maps.  In addition, we perform an external validation of our selection function using known X-ray clusters. 

The analytical model assumes pure Gaussian noise, in which case we simply have $P[q|\qm] = e^{-(q-\qm)^2/2}/\sqrt{2\pi}$.  The survey selection function is then given by the error function (and we refer to this as the ERF completeness function),
\begin{equation}
\label{eq:completeness}
\chi(\Yfive,\thetafive,l,b) = \frac{1}{2}\left[1-\mathrm{erf}\left(\frac{\qcat-\qm(\Yfive,\thetafive,l,b)}{\sqrt{2}}\right)\right].
\end{equation}
This model can be applied to a catalogue with well-defined noise properties (i.e., $\sigmaf$), such as our MMF3 catalogue, but not to the intersection catalogue based on the simultaneous detection with three different methods.  This is our motivation for choosing the MMF3 catalogue as our baseline.

In the Monte Carlo approach, we inject simulated clusters directly into the \Planck\ maps and (re)extract them with the complete detection pipeline.    Details are given in the accompanying 2015 SZ catalogue paper, \citet{planck2014-a36}.  This method provides a more comprehensive description of the survey selection by accounting for a variety of effects beyond noise.  In particular, we vary the shape of the SZ profile at fixed $\Yfive$ and $\thetafive$ to quantify its effect on catalogue completeness.  The difference between the Monte-Carlo and ERF completeness results in a change in modelled number counts of typically $\sim 2.5\%$ (with a maximum of $9\%$) in each redshift bin.

We also perform an external check of the survey completeness using known X-ray clusters from the Meta Catalogue of X-ray Clusters (MCXC) compilation \citep{2011A&A...534A.109P} and also SPT clusters from \citet{bleem2014}}.  Details are given in the 2015 SZ catalogue paper, \citet{planck2014-a36}.  For the MCXC compilation, we rely on the expectation that at redshifts $z<0.2$ any \Planck-detected cluster should be found in one of the ROSAT catalogues \citep{chamballu2012} because at low redshift ROSAT probes to lower masses than \Planck.\footnote{In fact, this expectation is violated to a small degree.  As discussed in~\cite{planck2014-a36}, there appears to be a small population of X-ray under-luminous clusters.} The MCXC catalogue provides a truth table, replacing the input cluster list of the simulations, and we compute completeness as the ratio of objects in the cosmology catalogue to the total number of clusters.
As discussed in \citet{planck2014-a36}, the results are consistent with Gaussian noise and bound the possible effect of profile variations.  We arrive at the same conclusion when applying the technique to the SPT catalogue.   

\citet{planck2014-a36} discusses completeness checks in greater detail.  One possible source of bias is the presence of correlated
IR emission from cluster member galaxies.  \citet{planck2014-a29} suggests that IR point sources may contribute significantly to the cluster SED at the \Planck\ frequencies, especially at higher redshift.  The potential impact of this effect warrants further study in future work. 

We thus have different estimations of the selection function for MMF3 and the intersection catalogues.  We test the sensitivity of our cosmological constraints to the selection function in the Appendix by comparing results obtained with the different methods and catalogues.  We find that our results are insensitive to the choice of completeness model (Fig~\ref{fig:erf_qa}), and we therefore adopt the analytical ERF completeness function for simplicity throughout the paper.

\section{The cluster mass scale}
\label{sec:cms}
The characteristic mass scale of our cluster sample is the critical element in our analysis of the counts.  It is controlled by the mass bias factor, $1-b$, accounting for any difference between the X-ray mass proxies used to establish the scaling relations and the true (halo) mass: $\Mx=(1-b)\Mfive$.  Such a difference could arise from cluster physics~\citep[such as a violation of hydrostatic equilibrium or temperature structure in the gas,][]{rasia2006, rasia2012, rasia2014}, from observational effects (e.g., instrumental calibration), or from selection effects biasing the X-ray samples relative to SZ- or mass-selected samples \citep{angulo2012}.

In our 2013 analysis, we adopted a flat prior on the mass bias over the range $1-b=[0.7,1.0]$, with a reference model defined by $1-b=0.8$. This was motivated by a comparison of the $Y-\Mx$ relation with published $Y-M$ relations derived from numerical simulations, as detailed in the Appendix of~\citet{planck2013-p15}; this estimate was consistent  with most (although not all) predictions for any violation of hydrostatic equilibrium, as well as observational constraints from the available lensing observations.  Effects other than cluster physics can contribute to the mass bias, as discussed in our earlier paper, and as emphasized by the survey of cluster multi-band scaling relations by \citet{rozo2014a,rozo2014b,rozo2014c}.

The mass bias was the largest uncertainty in our 2013 analysis, and it severely hampered understanding of the tension found between constraints from the primary CMB and the cluster counts.  Here, we incorporate new lensing mass determinations of \Planck\ clusters to constrain the mass bias.  We also apply a novel method to measure object masses based on lensing of CMB temperature anisotropies behind clusters \citep{melin2014}.  These constraints are used as prior information in our analysis of the counts.  As we will see, however, uncertainty in the mass bias remains our largest source of uncertainty, mainly because these various determinations continue to differ by up to 30\,\%.

In general, the mass bias could depend on cluster mass and redshift, although we will model it by a constant in the following.  Our motivation is one of practicality: the limited size and precision of current lensing samples makes it difficult to constrain any more than a constant value, i.e., the overall mass scale of our catalogue.  Large lensing surveys like \Euclid, WFIRST, and the Large Synoptic Survey Telescope, as well as CMB lensing, will improve this situation in coming years.

\subsection{Constraints from gravitational shear}
Several cluster samples with high quality gravitational shear mass measurements have appeared since 2013.  Among these, the Weighing the Giants \citep[WtG,][]{WtGI}, CLASH \citep{postman2012,merten2014,umetsu2014},
and Canadian Cluster Comparison Project \citep[CCCP,][]{hoekstra2015} programmes offer constraints on our mass bias factor, $1-b$, through direct comparison of the lensing masses to the \Planck\ mass proxy, $\MYsz$.  

The analysis by the WtG programme of 22 clusters from the 2013 \Planck\ cosmology sample yields $1-b=0.688\pm 0.072$.  
Their result lies at the very extreme of the range explored in \citet{planck2013-p15} and would substantially reduce the tension found between primary CMB and galaxy cluster constraints.  
\citet{hoekstra2015} report a smaller bias of $1-b=0.78 \pm 0.07\, {\rm (stat)}\,\pm 0.06\, {\rm (sys)}$ for a set of 20 common clusters, which is in good agreement with the fiducial value adopted in our 2013 analysis. In our new analysis we add the statistical and systematic uncertainties in quadrature (see Table~\ref{tab:mass_priors}).

The two samples overlap, but not completely, and, as discussed in detail by \citet{hoekstra2015}, there are numerous differences between the two analyses.  These include treatment of source redshifts, contamination by cluster members and methods of extracting a mass estimate from the lensing data.  And while the two mass calibrations differ in a way that attracts particular attention in the present context, they are statistically consistent, separated by about one standard deviation.

\subsection{Constraints from CMB lensing}

Measuring cluster mass through CMB lensing~\citep[][]{lewis2006} has been discussed in the literature for some time since the study performed by \citet{zaldarriaga1999}.  We apply a new technique for measuring cluster masses through lensing of CMB temperature anisotropies \citep{melin2014}, allowing us to calibrate the scaling relations using only \Planck\ data.  This is a valuable alternative to the galaxy lensing observations because it is independent and affected by different possible systematics.  Additionally, we can apply it to the entire cluster sample to obtain a mass calibration representative of an SZ flux-selected sample.  Similar approaches using CMB lensing to measure halo masses were recently applied by SPT~\citep{baxter2014} and ACT~\citep{madha2014}.

Our method first extracts a clean CMB temperature map with a constrained internal linear combination (ILC) of the \Planck\ frequency channels in the region around each cluster; the ILC is constrained to nullify the SZ signal from the clusters themselves and provide a clean CMB map of 5 arcmin resolution.  Using a quadratic estimator on the CMB map, we reconstruct the lensing potential in the field and then filter it to obtain an estimate of the cluster mass.  The filter is an NFW profile \citep{nfw} with scale radius set by the \Planck\ mass proxy for each cluster, and designed to return an estimate of the ratio $\Mlens/\MYsz$, where $\MYsz$ is the \Planck\ SZ mass proxy. These individual measurements are corrected for any mean-field bias by subtracting identical filter measurements on blank fields; this accounts for effects of apodization over the cluster fields and correlated noise.  The technique has been tested on realistic simulations of \Planck\ frequency maps.  More detail can be found in \citet{melin2014}.

\begin{figure}
\centering
\includegraphics[width=\hsize]{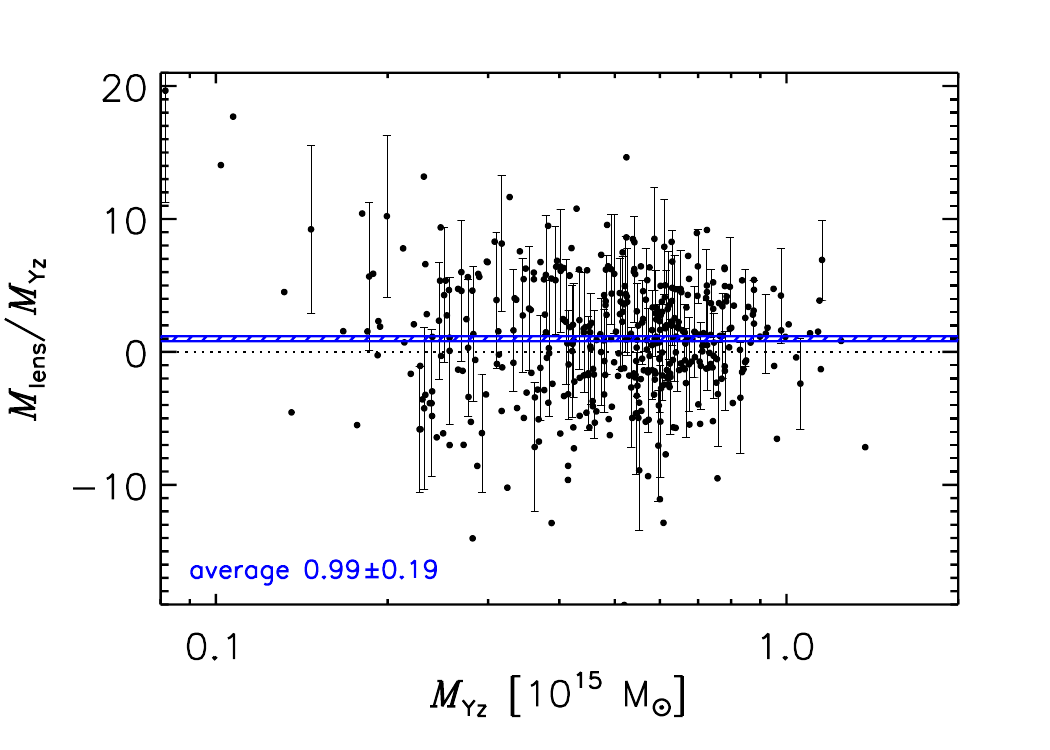}
\caption{Cluster mass scale determined by CMB lensing.  We show the ratio of cluster lensing mass, $\Mlens$, to the SZ mass proxy, $\MYsz$, as a function of the mass proxy for clusters in the MMF3 2015 cosmology sample.  The cluster mass is measured through lensing of CMB temperature anisotropies in the \Planck\ data \citep{melin2014}.  Individual mass measurements have low signal-to-noise, but we determine a mean ratio for the sample of $\Mlens/\MYsz = 1/(1-b) = 0.99\pm0.19$.  For clarity, only some of the error bars are plotted (see text).}
\label{fig:cmb_cms}
\end{figure}

Figure~\ref{fig:cmb_cms} shows $\Mlens/\MYsz$ as a function of $\MYsz$ for all clusters in the MMF3 cosmology sample.  Each point is an individual cluster.\footnote{The values can be negative due to noise fluctuations and the low signal-to-noise of  the individual  measurements.} For clarity, only some of the error bars on the ratio are shown; the error bars vary from $1.8$ at the high mass end to $8.5$ at the low mass end, with a median of 4.2.  There is no indication of a correlation between the ratio and $\MYsz$, and we therefore fit for a constant ratio of $\Mlens/\MYsz$ by taking the weighted mean (using the individual measurement uncertainties as provided by the filter) over the full data set.  If the ratio differs from unity, we apply a correction to account for the fact that our filter aperture was not perfectly matched to the clusters.  The correction is calculated assuming an NFW profile and is of the order of one percent.

The final result is $1/(1-b) = 0.99\pm 0.19$, traced by the blue band in the figure.  We note that the method constrains $1/(1-b)$ rather than $1-b$ as in the case of the shear measurements.  The calculated uncertainty on the weighted mean is consistent with a bootstrap analysis, where we create new catalogues of the same size as the original by sampling objects from the full catalogue with replacement; the uncertainty from the bootstrap is then taken as the standard deviation of the bootstrap means.   

The uncertainty 0.19 is statistical. \cite{melin2014} quote an uncertainty of 0.28 for the 62 ESZ-XMM clusters based on simulations including only tSZ, kSZ, primary CMB and instrumental noise. Scaling this number to 433 (number of objects with a redshift in the cosmological sample) gives $0.28 \,\sqrt{62/433} \approx 0.11$, in broad agreement with our value of 0.19. The difference can likely be attributed to the fact that the \Planck\ cosmological sample is on average less massive than the ESZ-XMM sample, and that the \Planck\ maps are more complex than the model adopted in~\cite{melin2014}. 

We have obtained a $5\sigma$ measurement of the sample mass scale using CMB lensing.  We emphasize, however, that the method is new and under development.  A number of potential systematic effects require further study, including cluster miscentring and mismatch between filter shape and actual cluster profiles; these would tend to reduce the observed masses from their true values.  On the other hand, effects such as contributions from mass correlated on large scales with the clusters (e.g., filaments or neighbouring halos) and contamination by infrared and radio sources could increase the observed signal.  We are examining these issues in a study of the ESZ-XMM sample that will be published at a later date.

\subsection{Summary}
The three mass bias priors are summarized in Table~\ref{tab:mass_priors}, and we will extract cosmological constraints from each one.  We favour these three lensing results because of their direct comparison to the \Planck\ mass proxy.  We will assume Gaussian distributions for $1-b$ (gravitational shear)  or $1/(1-b)$ (CMB lensing), with standard deviations given by the error column. We adopt the CCCP mass calibration as our baseline, and give the CMB lensing result less weight in our interpretation because of its novelty and the ongoing studies of the issues mentioned in the previous section.

\begin{table}[tb] 
\begingroup
\newdimen\tblskip \tblskip=5pt
\caption{Summary of mass scale priors}
  \label{tab:mass_priors}
\nointerlineskip
\vskip -3mm
\footnotesize
\setbox\tablebox=\vbox{
   \newdimen\digitwidth 
   \setbox0=\hbox{\rm 0} 
   \digitwidth=\wd0 
   \catcode`*=\active 
   \def*{\kern\digitwidth}
   \newdimen\signwidth 
   \setbox0=\hbox{+} 
   \signwidth=\wd0 
   \catcode`!=\active 
   \def!{\kern\signwidth}

\halign{#\hfil\tabskip=2em & \hfil#\tabskip=2em & \hfil#\tabskip=0pt\cr                           
\noalign{\doubleline}
Prior name & Quantity & Value and Gaussian errors\cr
\noalign{\vskip 3pt\hrule\vskip 2pt}
Weighing the Giants (WtG) & $ 1-b $  & $0.688 \pm 0.072$\cr
Canadian Cluster Comparison & &\cr
Project (CCCP) & $ 1-b $ & $0.780 \pm 0.092$\cr
CMB lensing (CMBlens)  & $1/(1-b)$ &$0.99*\pm 0.19$*\cr\hline\cr
Baseline 2013 & $ 1-b $ & $0.8 \, [-0.1,+0.2]$\cr
\noalign{\vskip 3pt\hrule\vskip 2pt}}}
\endPlancktable                    
\endgroup
\tablefoot{For CCCP, we use the value determined for \Planck\ clusters at S/N$>7$ \citep[][left column of p. 706]{hoekstra2015} and we add in quadrature the statistical (0.07) and systematic (0.06) uncertainties. CMB lensing directly measures $1/(1-b)$, which we implement in our analysis; purely for reference, this constraint translates approximately to $1-b=1.01^{+0.24}_{-0.16}$.  The last line shows the 2013 baseline --- a reference model defined by $1-b = 0.8$ with a flat prior in the [0.7, 1] range.
}
\end{table}                        

\section{Analysis methodology}
\label{sec:like}

\subsection{Likelihood}
\label{sec:likedef}
Our 2013 analysis employed a likelihood built on the cluster redshift distribution, $dN/dz$.  With the larger 2015 catalogue, our baseline likelihood is now constructed on counts in the $(z,q)$-plane.  We divide the catalogue into bins of size $\Delta z =0.1$ (10 bins) and $\Delta\log q =0.25$ (5 bins), each with an observed number $N(z_i,q_j)=N_{ij}$ of clusters.  Modelling the observed counts, $N_{ij}$, as independent Poisson random variables, our log-likelihood is
\begin{equation}
\label{eq:like2d}
\ln L = \sum_{i,j}^{N_z N_q} \left[ N_{ij}\ln\bar{N}_{ij}  - \bar{N}_{ij} - \ln[N_{ij}!]  \right],
\end{equation}
where $N_z$ and $N_q$ are the total number of redshift and signal-to-noise bins, respectively.  The mean number of objects in each bin is predicted by theory according to Eq.~(\ref{eq:dndzdq}):
\begin{equation}
\bar{N}_{ij} = \frac{dN}{dzdq}(z_i, q_j) \Delta z \Delta q,
\end{equation}
which depends on the cosmological (and cluster modelling) parameters.  In practice, we use a Monte Carlo Markov chain (MCMC) to map the likelihood surface around the maximum and establish confidence limits.

Eq.~(\ref{eq:like2d}) assumes that the bins are uncorrelated, while a more complete description would include correlations due to large-scale clustering.  In practice, our cluster  sample contains mostly high mass systems for which the impact of these effects is weak \citep[e.g.,][in particular their figure~4 for the impact on constraints in the ($\OmM$,$\sigma_8$) plane]{hu2003}.

\subsection{External data sets}
\label{sec:extdata}
Cluster counts cannot constrain all pertinent cosmological parameters; they are most sensitive to $\OmM$ and $\sigma_8$, and when analysing the counts alone we must apply additional observational constraints as priors on other parameters.  For this purpose, we adopt Big Bang nucleosynthesis (BBN) constraints from~\cite{steigman2008}, $\Omega_{\rm b} h^2 = 0.022 \pm 0.002$ \citep[for a recent review on BBN, see][]{olive2013}, and constraints from baryon acoustic oscillations (BAO).  The latter combine the 6dF Galaxy Survey~\citep{beutler2011}, the SDSS Main Galaxy Sample~\citep{padman2012,anderson2012} and the BOSS DR11~\citep{anderson2014}. We refer the reader to Sect. 5.2 in \citet{planck2014-a15} for details of the combination. We also include a prior on $n_{\rm s}$ from~\cite{planck2013-p11}, $n_{\rm s}=0.9624 \pm 0.014$.  When explicitly specified in the text, we add the supernov{\ae} constraint from SNLS-II and SNLS3: the Joint Light-curve Analysis constraint~\citep[JLA,][]{betoule2014}. The BAO are particularly sensitive to $H_0$, while the supernov{\ae} allow precise constraints on the dark energy equation-of-state parameter, $w$.

\section{Comparison to 2013}
\label{sec:comparison2013}
 
We begin by verifying  consistency with the results of \citet{planck2013-p15} (Sect.~\ref{sec:comp2013}) based on the one-dimensional likelihood over the redshift distribution, $dN/dz$ (Eq.~\ref{eq:dndz}).  We then examine the effect of changing to the full two-dimensional likelihood, $dN/dzdq$ (Eq.~\ref{eq:dndzdq}) in Sect.~\ref{sec:12Dcomp}.  For this purpose we compare constraints on the total matter density, $\OmM$, and the linear-theory amplitude of the density perturbations today, $\sigma_8$, using the cluster counts in combination with external data and fixing the mass bias. The two-dimensional likelihood $dN/dzdq$ is then adopted as the baseline in the rest of the paper.

\subsection{Constraints on $\OmM$ and $\sigma_8$: one-dimensional analysis}
\label{sec:comp2013}

\begin{figure}
\centering
\includegraphics[width=8cm]{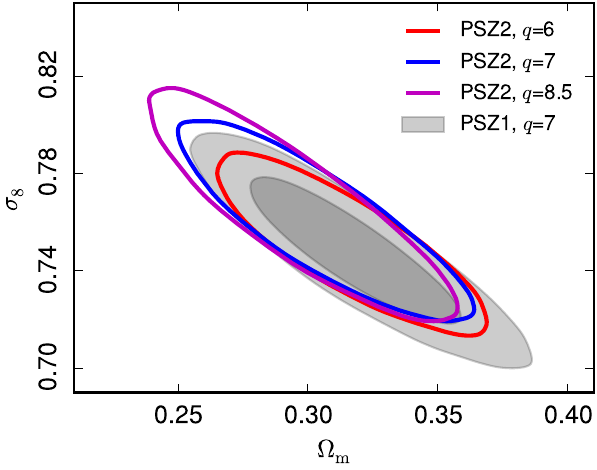}
\caption{Contours at 95\,\% for different signal-to-noise thresholds, $q=8.5$, 7, and 6, applied to the 2015 MMF3 cosmology sample for the SZ+BAO+BBN data set. The contours are compatible with the 2013 constraints \citep{planck2013-p15}, shown as the filled, light grey ellipses at 68 and 95\,\% (for the BAO and BBN priors of Sect~\ref{sec:extdata}; see text).   The 2015 catalogue thresholded at $q>8.5$ has a similar number of clusters (190) as the 2013 catalogue (189).  This comparison is made using the analytical error-function model for completeness and adopts the reference observable-mass scaling relation of the 2013 analysis ($1-b=0.8$, see text).  The redshift distributions of the best-fit models are shown in Fig.~\ref{fig:psz2_dndz}. For this figure and Fig.~\ref{fig:psz2_dndz}, we use the one-dimensional likelihood over the redshift distribution, $dN/dz$ (Eq.~\ref{eq:dndz}).
}
\label{fig:psz2_vs_threshold}
\end{figure}

Figure~\ref{fig:psz2_vs_threshold} presents constraints from the MMF3 cluster counts combined with the BAO and BBN priors of Sect.~\ref{sec:extdata};  we refer to this data combination as ``SZ+BAO+BBN''. To compare to results from our 2013 analysis (the grey, filled ellipses), we use a one-dimensional likelihood based on Eq.~(\ref{eq:dndz}) over the redshift distribution and have adopted the reference scaling relation of 2013, i.e., Eqs.~(\ref{eq:Yscaling}) and (\ref{eq:sizescaling}), with the mass bias value fixed to $1-b=0.8$.  For the present comparison, we use the updated BAO constraints discussed in Sect.~\ref{sec:extdata}; these are stronger than the BAO constraints used in the 2013 analysis, and the grey contours shown here are consequently smaller than in \citet{planck2013-p15}.

Limiting the 2015 catalogue to $q>8.5$ produces a sample with 190 clusters, similar to the 2013 cosmology catalogue (189 objects).  The two sets of constraints demonstrate good consistency, and they remain consistent while becoming tighter as we decrease the signal-to-noise threshold of the 2015 catalogue.  Under similar assumptions, our 2015 analysis thus confirms the 2013 results reported in \citet{planck2013-p15}.

The area of the ellipse from $q=8.5$ to $q=6$ decreases by a factor of 1.3. This is substantially less than the factor of 2.3 expected from the ratio of the number of objects in the two samples. The difference may be related to the decreasing goodness-of-fit of the best model as the signal-to-noise decreases. When incorporated, the uncertainty on the mass calibration $1-b$ will also restrict the reduction of the ellipse area.

\begin{figure}
\centering
\includegraphics[width=8cm]{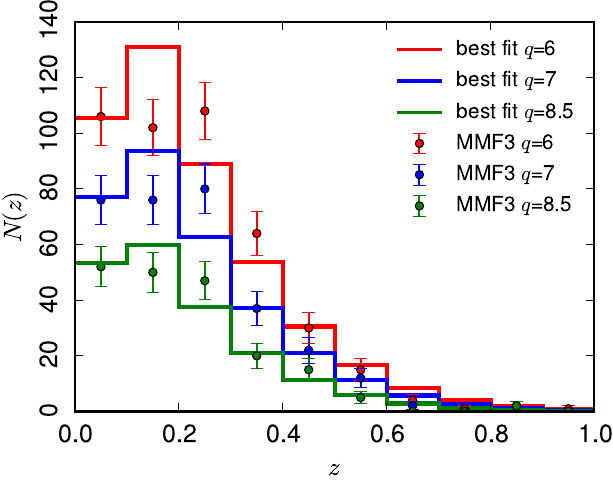}
\caption{Comparison of observed counts (points with error bars) with predictions of the best-fit models (solid lines) from the one-dimensional likelihood for three different thresholds applied to the 2015 MMF3 cosmology sample.  The mismatch between observed and predicted counts in the second and third lowest redshift bins, already noticed in the 2013 analysis, increases at lower thresholds, $q$.  The best-fit models are defined by the constraints shown in Fig.~\ref{fig:psz2_vs_threshold}. For this figure and Fig.~\ref{fig:psz2_vs_threshold}, we use our one-dimensional likelihood over the redshift distribution, $dN/dz$ (Eq.~\ref{eq:dndz}), with the mass biased fixed at $(1-b)=0.8$.
}
\label{fig:psz2_dndz}
\end{figure}

Figure~\ref{fig:psz2_dndz} overlays the observed cluster redshift distribution on the predictions from the best-fit model in each case.  We see that the models do not match the counts in the second and third redshift bins (counting from $z=0$), and that the discrepancy, already marginally present at the high signal-to-noise cut corresponding to the 2013 catalogue, becomes more pronounced towards the lower signal-to-noise thresholds.  This discrepancy cannot be attributed to redshift errors in the first bins because the majority of the redshifts are spectroscopic and the size of the bins is large ($\Delta_z=0.1$); for example, the first two redshift bins contain 208 clusters, of which 200 have spectroscopic redshifts.  The dependence on signal-to-noise may suggest that the data prefer a different slope, $\alpha$, of the scaling relation than allowed by the prior of Table~\ref{tab:params}.  
We explore the effect of relaxing the X-ray prior on $\alpha$ in the next section.

\subsection{Constraints on $\OmM$ and $\sigma_8$: two-dimensional analysis}
\label{sec:12Dcomp}

\begin{figure}
\centering
\includegraphics[width=9cm]{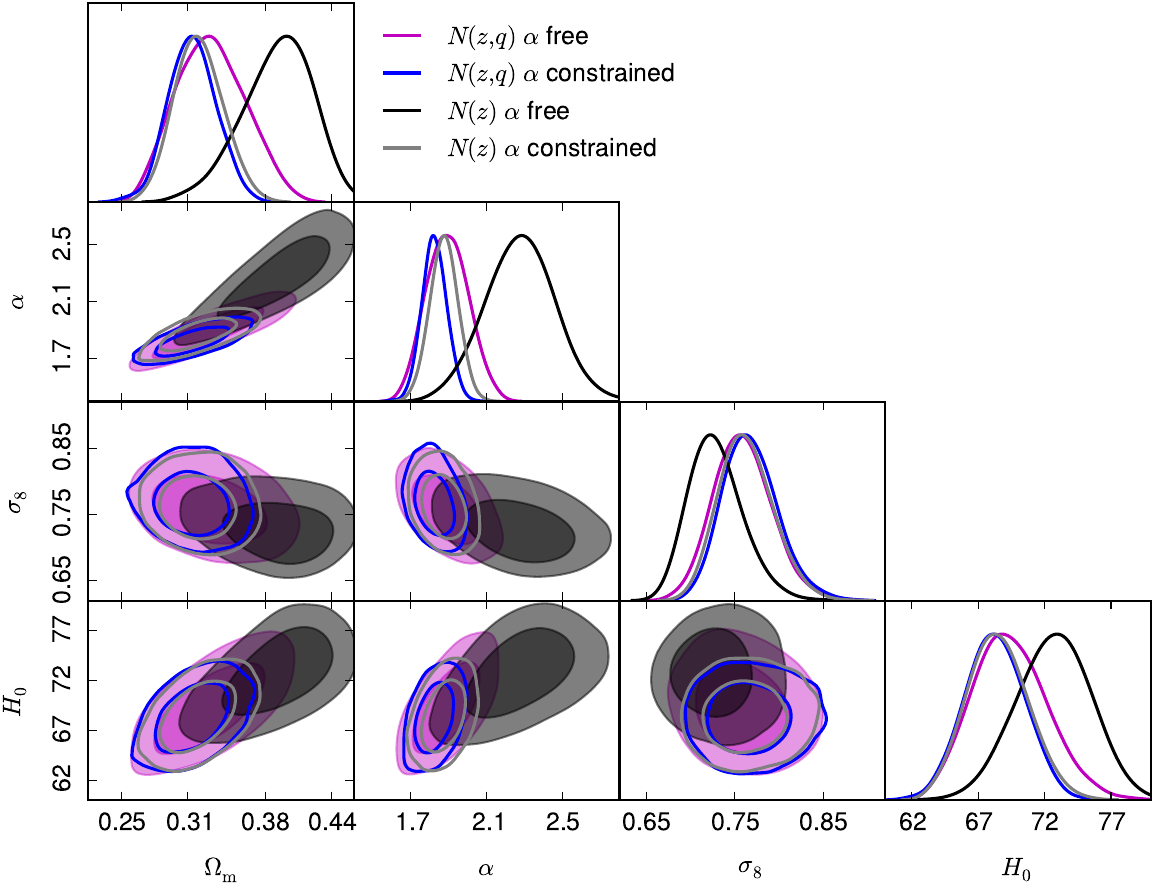}
\caption{Comparison of constraints from the one-dimensional ($dN/dz$) and two-dimensional ($dN/dzdq$) likelihoods on cosmological parameters and the scaling relation mass exponent, $\alpha$.  This comparison uses the MMF3 catalogue, the CCCP prior on the mass bias and the SZ+BAO+BBN data set.  The corresponding best-fit model redshift distributions are shown in Fig.~\ref{fig:psz2_dndz_comp}.
}
\label{fig:comp1d_2d}
\end{figure}

In Fig.~\ref{fig:comp1d_2d} we compare constraints from the one- and two-dimensional likelihood with $\alpha$ either free or with the prior of Table~\ref{tab:params}.  For this comparison, we continue with the ``SZ+BAO+BBN'' data set, but adopt the CCCP prior for the mass bias and only consider the full 2015 MMF3 catalogue at $q > 6$.

The grey and black contours and lines in Fig.~\ref{fig:comp1d_2d} show results from the one-dimensional likelihood fit to the redshift distribution using, respectively, the X-ray prior on $\alpha$ and leaving $\alpha$ free.  The redshift counts do indeed prefer a steeper slope, with a posterior of $\alpha= 2.23\pm 0.18$ in the latter case and a shift of the constraints along their degeneracy ridges.  To explore this preference, we split the analysis into low and high redshift bin sets divided at $z=0.2$, finding that neither the high nor the low redshift bin set prefers the steeper slope by itself; it appears only when analyzing all the bins.  As described further in the Appendix, there is a subtle interplay between parameters that is masked by the degeneracies and difficult to interpret with the present data set.

A related issue is the acceptability of the model fit.  We define a generalized $\chi^2$ measure of goodness-of-fit as $\chi^2 = \sum_{i}^{N_z} \bar{N}_{i}^{-1} \left( N_{i} - \bar{N}_{i} \right ) ^2$, determining the probability to exceed (${\rm PTE}$) the observed value using Monte Carlo simulations of Poisson statistics for each bin with the best-fit model mean $\bar{N}_{i}$.  The observed value of the fit drops from $17$ (${\rm PTE}=0.07$) with the X-ray prior, to $15$ (${\rm PTE}=0.11$) when leaving $\alpha$ free.  When leaving $\alpha$ free, $\OmM$ increases and $\sigma_8$ decreases, following their correlation with $\alpha$ shown by the contours, and their uncertainty increases due to the added parameter.    

The two-dimensional likelihood over $dN/dzdq$ better constrains the slope when $\alpha$ is free, as shown by the violet curves and contours.  In this case, the preferred value drops back towards the X-ray prior: $\alpha=1.89\pm 0.11$, just over $1\, \sigma$ from the central X-ray value.  Re-imposing the X-ray prior on $\alpha$ with the two-dimensional likelihood (blue curves) does little to change the parameter constraints.  Although the one-dimensional likelihood prefers a steeper slope than the X-ray prior, the two-dimensional analysis does not, and the cosmological constraints remain robust to varying $\alpha$.

We define a generalized $\chi^2$ statistic as described above, now over the two-dimensional bins in the $(z,q)$-plane.  This generalized $\chi^2$ for the fit with the X-ray prior is $43$ (${\rm PTE}=0.28$), compared to $\chi^2=45$ (${\rm PTE}=0.23$) when $\alpha$ is a free parameter.  

\begin{figure}
\centering
\includegraphics[width=8cm]{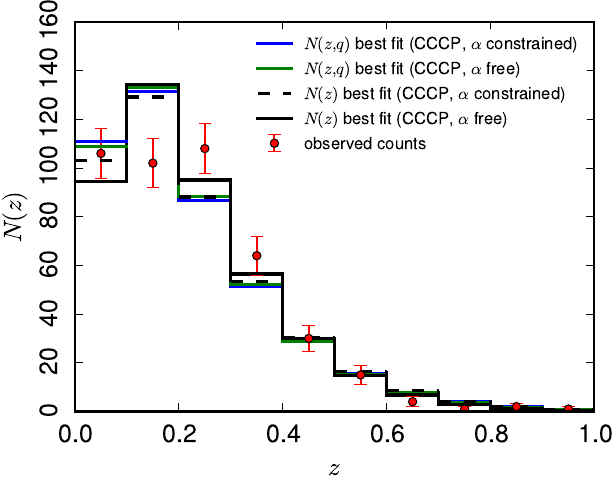}
\caption{Redshift distribution of best-fit models from the four analysis cases shown in Fig.~\ref{fig:comp1d_2d}.  The observed counts in the MMF3 catalogue ($q > 6$) are plotted as the red points with error bars, and as in Fig.~\ref{fig:comp1d_2d} we adopt the CCCP mass prior with the SZ+BAO+BBN data set.}
\label{fig:psz2_dndz_comp}
\end{figure}

Fig.~\ref{fig:psz2_dndz_comp} displays the redshift distribution of the best-fit models in all four cases.  Despite their apparent difficulty in matching the second and third redshift bins, the PTE values suggest that these fits are moderately good to acceptable. We note that, as mentioned briefly in Sect.~\ref{sec:likedef}, clustering effects will increase the scatter in each bin slightly over the Poisson value we have assumed, causing our quoted PTE values to be somewhat smaller than the true ones.

\section{Cosmological constraints 2015}
\label{sec:constraints}
We extract constraints on $\OmM$ and $\sigma_8$ from the cluster counts in combination with external data, imposing the different cluster mass scale calibrations as prior distributions on the mass bias.  In Sect.~\ref{sec:baselcdm}, we compare our new constraints to and then combine them with those from the CMB anisotropies in the base $\Lambda$CDM model.  We study parameter extensions to the base model in Sect.~\ref{sec:extensions}. In the following, we adopt as our baseline the 2015 two-dimensional SZ likelihood with the CCCP mass bias prior, $\alpha$ free and $\beta=2/3$ fixed in Eq.~(\ref{eq:Yscaling}). All quoted intervals are 68\% confidence and all upper/lower limits are 95\% confidence.

\subsection{Base $\Lambda$CDM}
\label{sec:baselcdm}

\begin{table*}[tb]
\begingroup
\newdimen\tblskip \tblskip=5pt
\caption{Summary of \Planck\ 2015 cluster cosmology constraints}
  \label{tab:cosmo_constraints}
\nointerlineskip
\vskip -3mm
\footnotesize
\setbox\tablebox=\vbox{
   \newdimen\digitwidth 
   \setbox0=\hbox{\rm 0} 
   \digitwidth=\wd0 
   \catcode`*=\active 
   \def*{\kern\digitwidth}
   \newdimen\signwidth 
   \setbox0=\hbox{+} 
   \signwidth=\wd0 
   \catcode`!=\active 
   \def!{\kern\signwidth}
\halign{#\hfil\tabskip=2em & \hfil#\tabskip=2em & \hfil#\tabskip=2em & \hfil#\tabskip=0pt\cr                           
\noalign{\doubleline}
Data & $\sigma_8 \left (\Omega_m \over 0.31 \right)^{0.3}$ &  $\Omega_m$ & $\sigma_8$\cr
\noalign{\vskip 3pt\hrule\vskip 1pt}
WtG + BAO + BBN & $0.806\pm 0.032$ & $0.34 \pm 0.03$ & $0.78 \pm 0.03$\cr
CCCP + BAO + BBN \bf{[Baseline]} & $0.774\pm 0.034$ & $0.33 \pm 0.03$ & $0.76 \pm 0.03$\cr
CMBlens + BAO + BBN & $0.723\pm 0.038$ & $0.32 \pm 0.03$ & $0.71 \pm 0.03$\cr
\noalign{\vskip 2pt\hrule\vskip 2pt}
CCCP + $H_0$ + BBN & $0.772\pm 0.034$ & $0.31 \pm 0.04$ & $0.78 \pm 0.04$\cr
\noalign{\vskip 1pt\hrule\vskip 1pt}}}
\endPlancktable                    
\endgroup
\tablefoot{The constraints are obtained for our baseline model:  the two-dimensional likelihood over the MMF3 catalogue ($q>6$) with $\alpha$ free and $\beta=2/3$ fixed in Eq.~(\ref{eq:Yscaling}).}
\end{table*}                        

\subsubsection{Constraints on $\OmM$ and $\sigma_8$: comparison to primary CMB parameters}
\label{sec:compCMB}
\begin{figure}
\centering
\includegraphics[width=8cm]{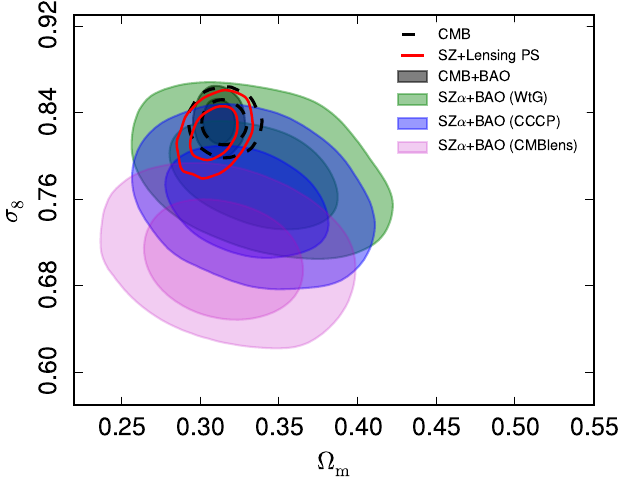}
\caption{Comparison of constraints from the CMB to those from the cluster counts in the ($\OmM, \sigma_8$)-plane.  The green, blue and violet contours give the cluster constraints (two-dimensional likelihood) at 68 and 95\,\% for the WtG, CCCP, and CMB lensing mass calibrations, respectively, as listed in Table~\ref{tab:mass_priors}.  These constraints are obtained from the MMF3 catalogue with the SZ+BAO+BBN data set and $\alpha$ free (hence the SZ$\alpha$ notation).  Constraints from the \Planck\ TT,\,TE,\,EE+lowP CMB likelihood (hereafter, \Planck\ primary CMB) are shown as the dashed contours enclosing 68 and 95\,\% confidence regions~\citep{planck2014-a15}, while the grey shaded region also includes BAO.  The red contours give results from a joint analysis of the cluster counts and the \Planck\ lensing power spectrum~\citep{planck2014-a17}, adopting our external priors on $n_s$ and $\Omega_bh^2$ with the mass bias parameter free and $\alpha$ constrained by the X-ray prior (hence the SZ notation without the subscript $\alpha$).   
}
\label{fig:psz2_CMB}
\end{figure}

Our 2013 analysis brought to light tension between constraints on $\OmM$ and $\sigma_8$ from the cluster counts and those from the primary CMB in the base $\Lambda$CDM model.  In that analysis, we adopted a flat prior on the mass bias over the range $1-b=[0.7,1.0]$, with a reference model defined by $1-b=0.8$ \citep[see discussion in the appendix of][]{planck2013-p15}.
Given the good consistency between the 2013 and 2015 cluster results (Fig.~\ref{fig:psz2_vs_threshold}), we expect the tension to remain under the same assumptions concerning the mass bias.  

Figure~\ref{fig:psz2_CMB} compares our 2015 cluster constraints (MMF3 SZ+BAO+BBN) to those for the base $\Lambda$CDM model from the \Planck\  CMB anisotropies.  The cluster constraints, given the three different priors on the mass bias, are shown by the filled contours at 68 and 95\,\% confidence, while the dashed black contours give the \Planck\ TT,TE,EE+lowP constraints \citep[hereafter \Planck\ primary CMB,][]{planck2014-a15}; the grey shaded regions add BAO to the CMB.   The central value of the WtG mass prior lies at the extreme end of the range used in 2013 (i.e., $1-b=0.7$); with its uncertainty range extending even lower, the tension with primary CMB is greatly reduced, as pointed out by \citet{vonderlinden2014}.  With similar uncertainty but a central value shifted to $1-b=0.78$, the CCCP mass prior results in greater tension with the primary CMB.  The lensing mass prior, finally, implies little bias and hence much greater tension.  

The red contours present results from a joint analysis of the cluster counts and the \Planck\ lensing power spectrum~\citep{planck2014-a17}, adopting our external priors on $n_s$ and $\Omega_bh^2$ with the mass bias parameter free and $\alpha$ constrained by the X-ray prior. It is interesting to note that these constraints are fully independent of those from the primary CMB, but are in good agreement with them, favouring only slightly lower values for $\sigma_8$.

Table~\ref{tab:cosmo_constraints} summarizes our cluster cosmology constraints for the base $\Lambda$CDM model for the different mass bias priors.  We give the  marginalized constraints on $\OmM$ and $\sigma_8$, as well as their combination that is most tightly constrained by the cluster counts.  In addition, in the last line we list constraints when replacing the BAO prior by a prior on $H_0$ from direct local measurements \citep{riess2011}: $H_0 = 73.8\pm 2.4$\,km s$^{-1}$ Mpc$^{-1}$.

\subsubsection{Joint \Planck\ 2015 primary CMB and cluster constraints}
\label{sec:joint}

\begin{figure}
\centering
\includegraphics[width=8cm]{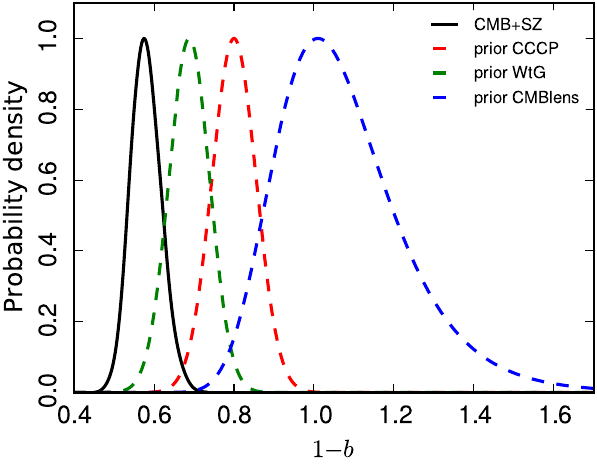} 
\caption{Comparison of cluster and primary CMB constraints in the base $\Lambda$CDM model, expressed in terms of the mass bias, \mbox{$1-b$}. The solid black curve shows the distribution of values required to reconcile the counts and primary CMB in $\Lambda$CDM; it is found as the posterior on $1-b$ from a joint analysis of the \Planck\ cluster counts and primary CMB when leaving the mass bias free.  The coloured dashed curves show the three prior distributions on the mass bias listed in Table~\ref{tab:mass_priors}.}
\label{fig:psz2_posterior_priors}
\end{figure}

\paragraph{Mass bias required by the primary CMB}
\label{sec:joint1}
In Fig.~\ref{fig:psz2_posterior_priors} we compare the three prior distributions to the mass bias required by the primary CMB.  The latter is obtained as the posterior on $1-b$ from a joint analysis of the MMF3 cluster counts and the CMB with the mass bias as a free parameter.  The best-fit value in this case is $1-b=0.58\pm 0.04$, more than 1$\sigma$ below the central WtG value.  Perfect agreement with the primary CMB would imply that clusters are even more massive than the WtG calibration.  This figure most clearly quantifies the tension between the \Planck\ cluster counts and primary CMB.  

\paragraph{Reionization optical depth}
Primary CMB temperature anisotropies also provide a precise measurement of the parameter combination $\As e^{-2\tau}$, where $\tau$ is the optical depth from Thomson scatter after reionization and $\As$ is the power spectrum normalization on large scales \citep[][]{planck2014-a15}.  Low-$\ell$ polarization anisotropies break the degeneracy by constraining $\tau$ itself, but this measurement is delicate given the low signal amplitude and difficult systematic effects; it is important, however, in the determination of $\sigma_8$. 
It is therefore interesting to compare the \Planck\ primary CMB constraints on $\tau$ to those from a joint analysis of the cluster counts and primary CMB without the low-$\ell$ polarization data (lowP).  \citet{battye2014}, for instance, pointed out that a lower value for $\tau$ than suggested by WMAP could reduce the level of tension between CMB and large-scale structure.

\begin{figure}
\centering
\includegraphics[width=8cm]{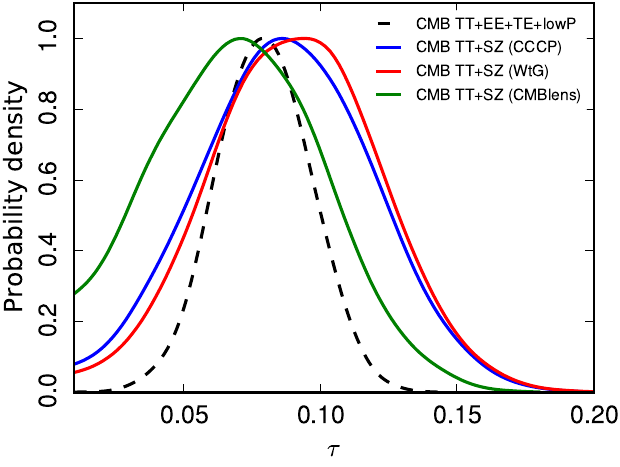}
\caption{Constraints on the reionization optical depth, $\tau$.  The dashed black curve is the constraint from \Planck\ CMB (i.e., TT,TE,EE+lowP), while the three coloured lines are the posterior distribution on $\tau$ from a joint analysis of the cluster counts and \Planck\ TT only for the three different mass bias parameters.
}
\label{fig:psz2_tau}
\end{figure}

The comparison is shown in Fig.~\ref{fig:psz2_tau}.  We see that the \Planck\ TT + SZ constraints are in good agreement with the value from \Planck\ CMB (i.e., TT,TE,EE+lowP), with the preferred value for WtG slightly higher and CMB lensing pushing towards a lower value. The ordering CMB lensing/CCCP/WtG from lower to higher $\tau$ posterior values matches the decreasing level of tension with the primary CMB on $\sigma_8$.  These values remain, however, larger than what is required to fully remove the tension in each case. The posterior distributions for the mass bias are $1-b=0.60 \pm 0.042$, $1-b=0.61 \pm 0.049$, $1-b=0.66 \pm 0.045$, respectively, for WtG, CCCP and CMB lensing, all significantly shifted from the corresponding priors of Table~\ref{tab:mass_priors}.  Allowing $\tau$ to adjust offers only minor improvement in the tension reflected by Fig.~\ref{fig:psz2_posterior_priors}.  Interestingly, the \Planck\ TT posterior shown in Fig. 8 of~\citet{planck2014-a15} peaks at significantly higher values, while our \Planck\ TT + SZ constraints are consistent with the result from \Planck\ TT + lensing, an independent constraint on $\tau$ without lowP.

\subsection{Model extensions}
\label{sec:extensions}

\subsubsection{Curvature}
We consider constraints on spatial curvature that can be set by cluster counts.  Our cluster counts combined with BBN and BAO for the CCCP mass prior yield $\Omega_K = -0.06\pm 0.06$.  This is completely independent of the CMB, but consistent with the CMB plus BAO constraint of $\Omega_K = 0.000\pm 0.002$.

\subsubsection{Dark energy}

\begin{figure}
\centering
\includegraphics[width=8cm]{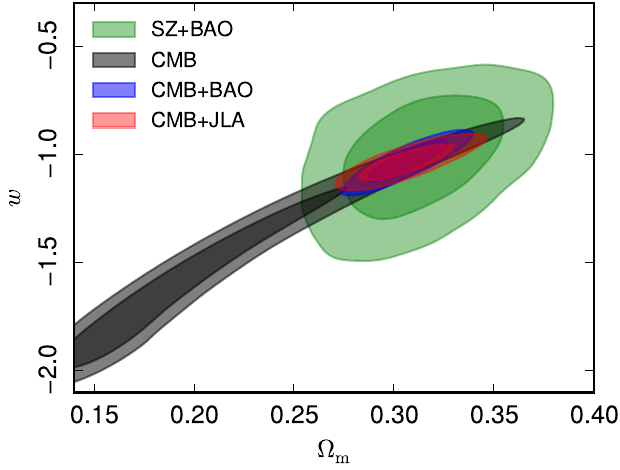}
\caption{Constraints on a constant dark energy equation-of-state parameter, $w$.  Analysis of the primary CMB alone yields the grey contours that are highly degenerate.  Adding either BAO or supernovae to the CMB breaks the degeneracy, giving constraints around $w=-1$.  The green contours are constraints from joint analysis of the SZ counts and BAO; although much less constraining they agree with the CMB+JLA combinations and are completely independent.}
\label{fig:psz2_w}
\end{figure}

Constraints on dark energy and modified gravity based on \Planck\ CMB and external data sets are studied in detail in~\cite{planck2014-a16}. In Fig.~\ref{fig:psz2_w} we examine constraints on a constant dark energy equation-of-state parameter, $w$.  Analysis of the primary CMB alone results in the highly degenerate grey contours.  The degeneracy is broken by adding constraints such as BAO (blue contours) or supernovae distances (rose-colored contours), both picking values around $w=-1$.
The SZ counts (two-dimensional likelihood with CCCP prior) only marginally break the degeneracy when combined with the CMB,
but when combined with BAO they do yield interesting constraints (green contours) that are consistent with the independent constraints from the primary CMB combined with supernovae. We obtain $\OmM= 0.314 \pm 0.026 $ and $w= -1.01 \pm 0.18 $ for SZ+BAO, $\OmM= 0.306 \pm 0.013 $ and $w= -1.10 \pm 0.06 $ for CMB+BAO, and $\OmM= 0.306 \pm 0.015 $ and $w= -1.10 \pm 0.05 $ for CMB+JLA.

\begin{figure*}
\centering
\includegraphics[width=15cm]{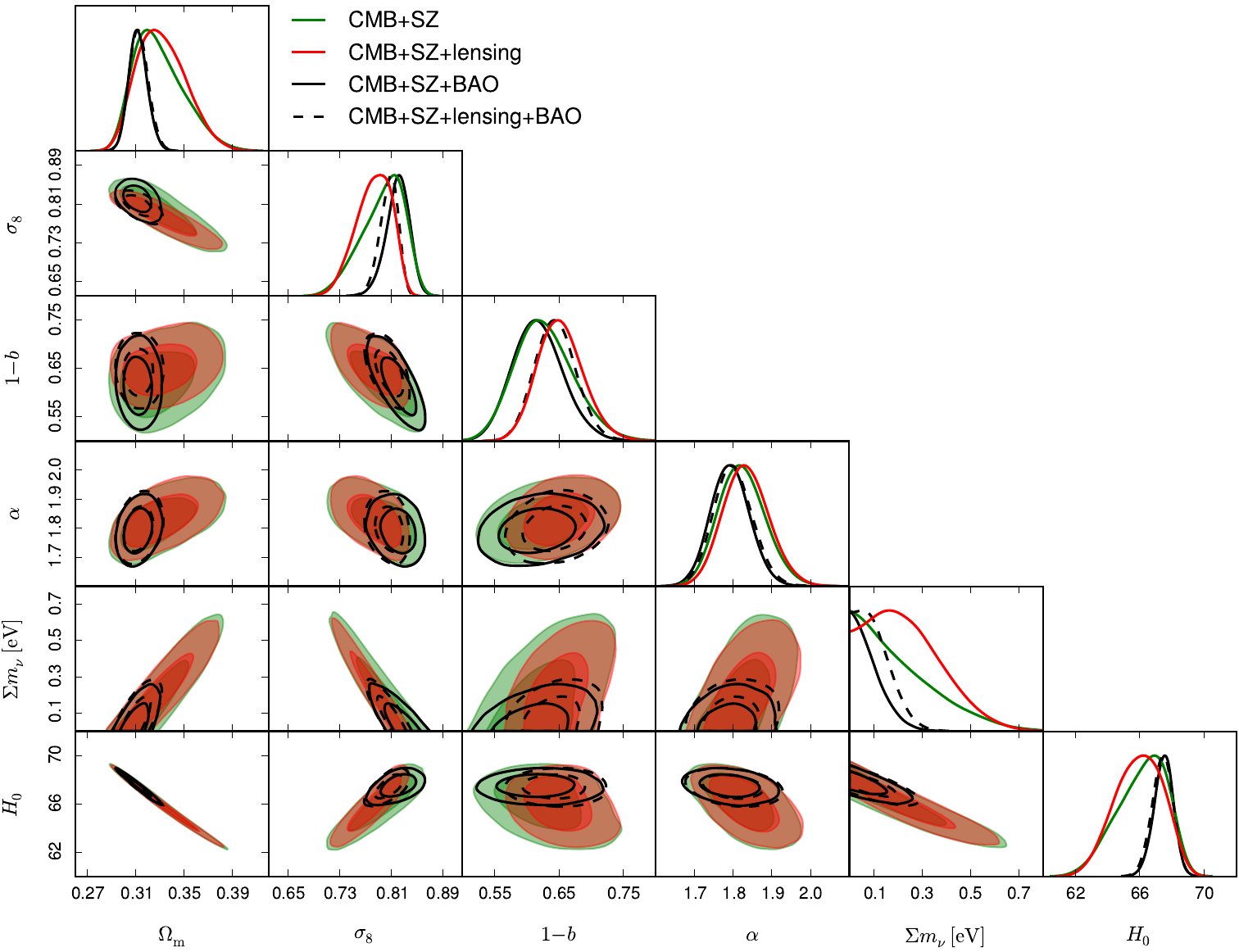}
\caption{Parameter constraints on the $\Lambda$CDM+non-minimal neutrino mass model.  For this study, we adopt the CCCP prior on the mass bias (see Table~\ref{tab:mass_priors}) and leave the scaling exponent, $\alpha$, free.  The green and red shaded regions show, respectively, the 68 and 95\,\% confidence regions for joint analyses of the cluster counts using the primary CMB, and the primary CMB plus the lensing power spectrum.  The solid and dashed black contours add to these two cases constraints from BAO. }
\label{fig:psz2_cccp_all}
\end{figure*}

\subsubsection{$\sum m_\nu$}
An important, well-motivated extension to the base $\Lambda$CDM model that clusters can help constrain is a non-minimal sum of neutrino masses, $\sum m_\nu > 0.06$\, eV.  Given the primary CMB anisotropies, the amplitude of the density perturbations today, characterized by the equivalent linear theory extrapolation, $\sigma_8$, is model dependent; it is a derived parameter, depending, for example, on the composition of the matter content of the Universe.  Cluster abundance, on the other hand, provides a direct measurement of $\sigma_8$ at low redshifts, and comparison to the value derived from the CMB tests the adopted cosmological model.   

By free-streaming, neutrinos damp the growth of matter perturbations.  Our discussion thus far has assumed the minimum mass for the three known neutrino species.  Increasing their mass, $\sum m_\nu > 0.06$\, eV, lowers $\sigma_8$ because the neutrinos have larger gravitational influence on the total matter perturbations.  This goes in the direction of reconciling tension --- the strength of which depends on the mass bias --- between the cluster and primary CMB constraints.  Cluster abundance, or any measure of $\sigma_8$ at low redshift, is therefore an important cosmological constraint to be combined with those from the primary CMB.  

Figure~\ref{fig:psz2_cccp_all} presents a joint analysis of the cluster counts for the CCCP mass bias prior with primary CMB, the \Planck\ lensing power spectrum, and BAO. 
The results without BAO (green and red shaded contours) allow relatively large neutrino masses, up to $\sum m_\nu \approx 0.5$\,eV; and when adding the lensing power spectrum, a small, broad peak appears in the posterior distribution just above $\sum m_\nu=0.2$\,eV.  We also notice some interesting correlations: the amplitude, $\sigma_8$, anti-correlates with neutrino mass, as does the Hubble parameter, and larger values of $\alpha$ correspond to larger neutrino mass, lower $\Ho$, and lower $\sigma_8$.

As discussed in detail in \citet{planck2014-a15}, the anti-correlation with the Hubble parameter maintains the observed acoustic peak scale in the primary CMB.  Increasing neutrino mass to simultaneously accommodate the cluster and primary CMB constraints by lowering $\sigma_8$, while allowed in this joint analysis, would therefore necessarily increase tension with most direct measurements of $\Ho$ \citep[see discussion in][]{planck2014-a15}.  Including the BAO data greatly restricts this possibility, as shown by the solid and dashed black curves.  

\begin{figure}
\centering
\includegraphics[width=8.5cm]{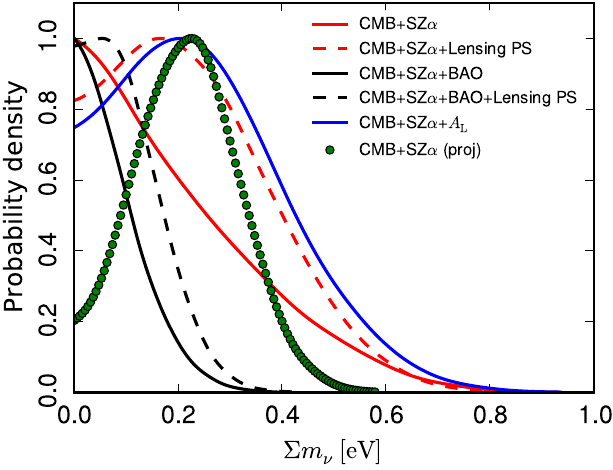}
\caption{Constraints on $\sum m_\nu$ from a joint analysis of the cluster counts and primary CMB.  The solid and dashed, red and black lines reproduce the marginalized posterior distributions from Fig.~\ref{fig:psz2_cccp_all}.  The solid blue line is the posterior from a similar analysis, but marginalized over the additional parameter $A_{\rm L}$ (see text).  If applied to the present \Planck\ cluster cosmology sample, a future mass calibration of $1-b=0.78 \pm0.01$ would result in the bold, dotted green posterior curve.}
\label{fig:sz_cmb_mnu}
\end{figure}

The solid and dashed, red and black curves in Fig.~\ref{fig:sz_cmb_mnu} reproduce the marginalized posterior distributions on $\sum m_\nu$ from Fig.~\ref{fig:psz2_cccp_all}.  The solid blue curve is the result of a similar analysis (CMB+SZ) where, in addition, the artificial parameter $A_{\rm L}$ is allowed to vary. This parameter characterizes the amount of lensing in the temperature power spectrum relative to the best fit model \citep{planck2014-a15}.
\Planck\ TT+lowP alone constrains
\begin{equation*}
A_{\rm L} = 1.22 \pm 0.10,
\end{equation*}
which is in mild tension with the value predicted for the $\Lambda$CDM model, $A_{\rm L} = 1$.
In the base $\Lambda$CDM model, this parameter is fixed to unity, but it is important to note that it is degenerate with $\sum m_\nu$.  Left free, it allows less lensing power, which is also in line with the direct measurement of the lensing power spectrum (labelled as ``Lensing PS'') from the four-point function \citep[see][]{planck2014-a15}.  In that light, we see that adding $A_{\rm L}$ as a free parameter accentuates the peak in the CMB+SZ+Lensing PS posterior. The small internal tension between CMB+SZ and CMB+SZ+$A_{\rm L}$ posteriors may point towards a need for an extension of the minimal six-parameter $\Lambda$CDM.

These posteriors lead to the following constraints: 
\begin{eqnarray}
\sum m_\nu & < & 0.50 \, {\rm eV\ (95\%)\ for\ CMB+SZ+Lensing PS}\\
\sum m_\nu & < & 0.20 \, {\rm eV\ (95\%)\ for\ CMB+SZ+BAO.}
\end{eqnarray}

We may compare these with the constraints from the primary CMB presented in \citet{planck2014-a15}.  The \Planck\ primary CMB by itself places an upper limit of $\sum m_\nu <0.49$\,eV (95\%), and the addition of BAO tightens this to $\sum m_\nu <0.17$\,eV (95\%).  Addition of the \Planck\ lensing power spectrum to the primary CMB weakens the constraint to $\sum m_\nu <0.59$\,eV (95\%), as we would expect given the results and discussion above.  The final constraint adopted by \citet{planck2014-a15}, for its robustness to possible remaining low level systematics in the polarization data, is $\sum m_\nu <0.23$\,eV (95\%), not too different from the peak suggested in CMB+SZ+lensing PS posterior.  

Adding neutrino mass should lower $\sigma_8$, letting it move towards values favoured by the cluster counts.  We might  expect that the CMB+SZ combination would therefore favour non-minimal neutrino mass.  In spite of this, the green curve only places an upper limit on $\sum m_\nu$.  We may understand this by looking at the posterior on the mass bias $1-b$ in Fig.~\ref{fig:psz2_cccp_all}  The allowed values are well separated from the prior distribution (CCCP), meaning that the primary CMB has sufficient statistical weight to strongly override the prior.  The lensing power spectrum, in favouring slightly lower $\sigma_8$, reinforces the cluster trend, so that a peak appears in the posterior for $\sum m_\nu$ in the red curve; it is not enough, however, to bring the posterior on the mass bias in line with the prior.  This indicates that the tension between the cluster and primary CMB constraints is not fully resolved.  

\begin{figure}
\centering
\includegraphics[width=8cm]{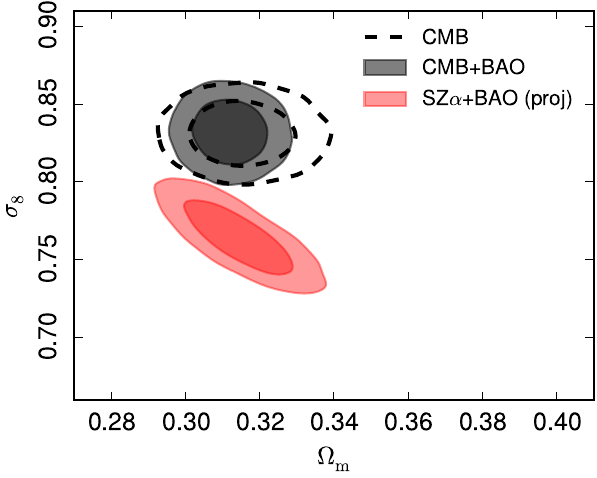}
\caption{Prediction of cluster constraints with a possible future mass bias prior of $1-b=0.78 \pm 0.01$.  The black shaded region and dashed contours reproduce the current primary CMB and primary CMB+BAO constraints from \Planck\ for the base $\Lambda$CDM model.  The red shaded contours present the constraints expected from this mass bias prior applied to the present \Planck\ cluster cosmology sample with the SZ+BAO+BBN data set.}
\label{fig:sz_oms8_proj}
\end{figure}

One may then ask, how tight must the prior on the mass bias be to make a difference?  To address this question, we performed an analysis assuming a projected tighter prior constraint on the mass bias.  The informal target precision for cluster mass calibration with future large lensing surveys, such as \Euclid\ and the Large Synoptic Survey Telescope, is 1\,\%. We therefore consider the impact of a prior of $1-b=0.78\pm0.01$ on the present \Planck\ cluster cosmology sample in Figs.~\ref{fig:sz_cmb_mnu} and \ref{fig:sz_oms8_proj}.  

The latter figure compares the constraints from cluster counts for this mass bias to the present primary CMB constraints in the $(\OmM, \sigma_8)$-plane for the base $\Lambda$CDM model.  The bold, green dotted curve in Fig.~\ref{fig:sz_cmb_mnu} shows the predicted posterior on the neutrino mass from a joint analysis of the present \Planck\ cluster counts and primary CMB with this projected mass bias prior.  The same prior on a much larger catalogue would demonstrate a corresponding increase in sensitivity to neutrino mass. This simple projection highlights the importance and value of the more precise cluster cosmology expected in the future, and it provides clear motivation for further significant effort in mass calibration.  This effort will continue with larger samples of clusters with gravitational shear measurements, and also with the new technique of CMB lensing cluster mass measurements.

In general, the mass bias is expected to depend on both mass and redshift. A more precise mass calibration study based on much larger samples, e.g., \Euclid\ or LSST, would be required to determine functional forms for the mass bias.  A dependence on mass/redshift can be accounted for by appropriate values for $\alpha$/$\beta$ in our formalism.

\section{Summary and discussion}
\label{sec:discussion}

Our 2015 analysis incorporates a number of improvements and new information relative to our first study in \cite{planck2013-p15}.  With more data, we have a larger cluster cosmology sample, increased by more than a factor of 2, and we implement a two-dimensional likelihood over the counts in both redshift and signal-to-noise.  We have also performed new tests of the selection function using MCXC and SPT cluster catalogues as truth tables.  The selection function from these external checks and internal simulations of the \Planck\ catalogue construction agree with each other and can be reasonably modelled by a simple analytical expression derived by assuming noise is the dominant factor (see the Appendix).  One possible systematic effect that warrants further study is IR emission from cluster member galaxies.  Finally, we have examined the implications of three recent determinations of the cluster mass bias parameter, $1-b$.  The two-dimensional likelihood with the 2015 catalogue and mass bias priors will be implemented in {\tt CosmoMC}.

Our analysis confirms the results of the 2013 study.  The counts are consistent with those of 2013, illustrated by the agreement in the constraints on $\OmM$ and $\sigma_8$ when using the same SZ observable-mass relations (see Fig.~\ref{fig:psz2_vs_threshold}).  The gain in statistical precision is less than expected from the larger catalogue, which is likely related to the fact that the fit to the redshift distribution with the X-ray prior on $\alpha$ is only marginal.  Our new two-dimensional approach yields consistent but more robust constraints than the one-dimensional likelihood over just the redshift distribution; it is less sensitive to the slope of the scaling relation, $\alpha$, and it provides a better fit to the counts than in the one-dimensional case.   

Using the two-dimensional likelihood as our baseline, we extracted new cosmological constraints using three different cluster mass scales represented by the mass bias prior distributions given in Table~\ref{tab:mass_priors}.  The first two come from galaxy shear observations of samples of \Planck\ clusters.  They differ by about $1\sigma$, with the WtG result favouring larger mass bias.  We have also implemented a novel method for measuring cluster masses based on lensing of the CMB temperature anisotropies behind clusters \citep{melin2014}; it gives a mass bias averaged over the entire cluster cosmology sample, although with larger statistical uncertainty.  

As a new method requiring further exploration, we consider CMB lensing less robust at present than galaxy lensing mass measurements, but highly promising.  Similar CMB-based mass measurements have recently been published by SPT \citep{baxter2014} and ACT \citep{madha2014}.  The approach is appealing because it is subject to different systematic effects than gravitational shear and because it can be applied to large cluster samples, thanks to the extensive sky coverage of the CMB experiments, with \Planck\ of course covering the entire sky.  Gravitational shear surveys will attain large sky coverage in the near future with the Dark Energy Survey (DES), and in the more distant future with the \Euclid\ and WFIRST space missions and the Large Synoptic Survey Telescope. 

Our central result from analysis of the 2015 \Planck\ cluster counts is shown in Fig.~\ref{fig:psz2_CMB}. Depending on the mass bias prior, we find varying degrees of tension with the primary CMB results, as in 2013.  The mass bias required to bring the cluster counts and CMB into full agreement is larger than indicated by any of the three priors and corresponds to $1-b=0.58\pm 0.04$.  Fig.~\ref{fig:psz2_posterior_priors} illustrates the situation.  The WtG prior almost eliminates the tension, but not quite, while both the CCCP and CMB lensing priors remain in noticeable disagreement. Our largest source of modelling uncertain is, as in 2013, the mass bias.

Tension between low redshift determinations of $\sigma_8$ and the \Planck\ primary CMB results are not unique to the \Planck\ cluster counts.  Among SZ cluster surveys, both SPT and ACT are in broad agreement with our findings, the latter depending on which SZ-mass scaling relation is used, as detailed in our 2013 analysis \citep{planck2013-p15}.  Furthermore the new SPT cosmological analysis~\citep{bocquet2014} shows a significant shift between the cluster mass scale determined from the velocity dispersion or $Y_{\rm X}$ and what is needed to satisfy \Planck\ or WMAP9 CMB constraints \citep[e.g., figure~2][]{bocquet2014}.  In a study of the REFLEX X-ray luminosity function, \citet{boehringer2014} also report general agreement with our cluster findings.  On the other hand, \citet{WtGIV} find that their X-ray cluster counts, when using the WtG mass calibration, match the primary CMB constraints. \cite{angrick2015} also find good agreement with the primary CMB constraints, fitting their X-ray temperature function with results from hydrodynamical simulations, and \cite{simet2015} recently measured $<M_{\rm X}/M_{\rm WL}>=0.66^{+0.07}_{-0.12}$ for the RBC X-ray galaxy cluster catalogue.

The situation is thus not yet satisfactory.  It is unclear if these modest tensions arise from low-level systematics in the astrophysical studies, or are the first glimpse of something more important.  The most obvious extension to the base $\Lambda$CDM model that could in principle reconcile the differences is a non-minimal sum of neutrino masses.  This, unfortunately, does not provide such a straightforward solution.  While it is true that adding neutrino mass does lower $\sigma_8$ relative to the base $\Lambda$CDM prediction from the primary CMB, it does so at the the cost of increasing tension in other parameters; for example, it lowers \Planck's value for the Hubble constant which is already lower than many direct estimates.  

\begin{figure}
\centering
\includegraphics[width=9cm]{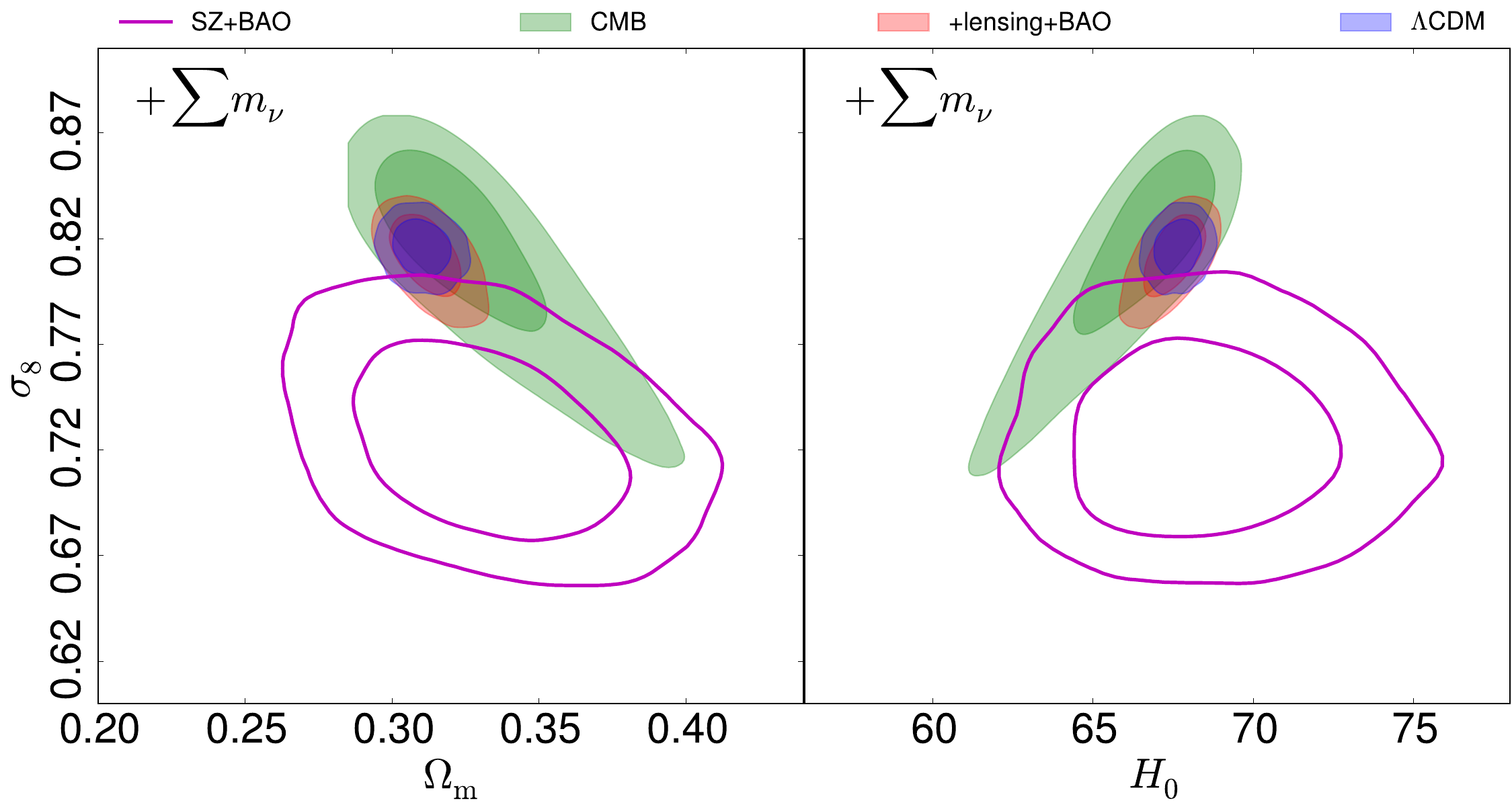}
\caption{Effects of neutrino mass. The open solid magenta contours reproduce our cluster constraints when marginalizing over variable neutrino mass. The violet shaded contours trace the constraints on the base $\Lambda$CDM model (with fixed neutrino mass), while the other shaded regions give constraints from the primary CMB only or combined with lensing and BAO when adding and marginalizing over variable neutrino mass. In this figure, the CMB likelihood is based on  \Planck\ TT,\,TE,\,EE+lowP while only \Planck\ TT+lowP is used in the equivalent figure in~\cite{planck2014-a15}.
}
\label{fig:comparisonwithprimarycmbpaper}
\end{figure}

Figure~\ref{fig:comparisonwithprimarycmbpaper} highlights these points by showing constraints in the $(\OmM,\sigma_8)$- and $(\Ho,\sigma_8)$-planes for the CCCP mass bias parameter.  Adding variable neutrino mass relaxes constraints from the primary CMB (shaded contours) towards lower $\sigma_8$, but while simultaneously increasing $\OmM$ and decreasing $\Ho$.
And the tension remains pronounced, regardless of the neutrino mass.

Another possibility is that baryonic physics influences the late-time evolution of the density perturbations.  Strong feedback from active galactic nuclei~\citep{vandaalen2011,martizzi2014} can potentially damp growth and lower $\sigma_8$ through expulsion of matter from dark matter halos.  This same effect could also reduce the mass of cluster halos and hence the prediction for their abundance, which is based on dark matter only simulations \citep{cui2014,velliscig2014,bocquet2015}.  It does not appear, however, that these effects are sufficiently large to explain the tension between low redshift and primary CMB constraints hinted at by the different observations \citep[e.g.,][]{cusworth2014,costanzi2014}.
In addition, the violent feedback necessary for important impact might be difficult to reconcile with observations of the baryon content of dark matter halos \citep[e.g.,][]{planck2012-XI, greco2015}, altough this point is still under discussion~\citep[e.g.,][]{planelles2014}.

In conclusion, we return to the main uncertainty in interpretation of the cluster counts, namely the mass bias.  It could be argued that the current accuracy is at the level of 10 - 15\,\%, based on the difference between different analyses and somewhat larger than their quoted statistical uncertainties.  Progress will certainly follow with improvement in these measurements.  We illustrate the potential impact of a 1\% determination of the mass bias in Figs.~\ref{fig:sz_cmb_mnu} and \ref{fig:sz_oms8_proj}.  Such a result would, depending on the central value, significantly clarify the extent of any tensions and possible necessity for extensions to the base $\Lambda$CDM model.  This precision is the avowed target of the large lensing surveys, such as \Euclid, WFIRST, and LSST.  In the shorter term, we may expect valuable movement in this direction from DES and CMB lensing cluster mass measurements.    

\begin{acknowledgements}
The Planck Collaboration acknowledges the support of: ESA; CNES, and CNRS/INSU-IN2P3-INP (France); ASI, CNR, and INAF (Italy); NASA and DoE (USA); STFC and UKSA (UK); CSIC, MINECO, JA and RES (Spain); Tekes, AoF, and CSC (Finland); DLR and MPG (Germany); CSA (Canada); DTU Space (Denmark); SER/SSO (Switzerland); RCN (Norway); SFI (Ireland); FCT/MCTES (Portugal); ERC and PRACE (EU). A description of the Planck Collaboration and a list of its members, indicating which technical or scientific activities they have been involved in, can be found at \href{http://www.cosmos.esa.int/web/planck/planck-collaboration}{http://www.cosmos.esa.int/web/planck/planck-collaboration}.
\end{acknowledgements}


\bibliographystyle{aa}
\bibliography{szcosmo2014,Planck_bib}


\appendix
\section{Modelling uncertainties}
\label{sec:robustness}
We examine the robustness of our cosmological constraints to modelling uncertainties.  We first consider sensitivity to the cosmological sample and to our modelling of the completeness function in Sect.~\ref{sec:rob_selfun}, and then look at the effect of using a different mass function in Sec.~\ref{sec:rob_mf}.  In Sec.~\ref{sec:rob_evol}, we show that our constraints are robust against redshift evolution of the scaling relations.

\subsection{Choice of the sample and selection function}
\label{sec:rob_selfun}
\begin{figure}
\centering
\includegraphics[width=8cm]{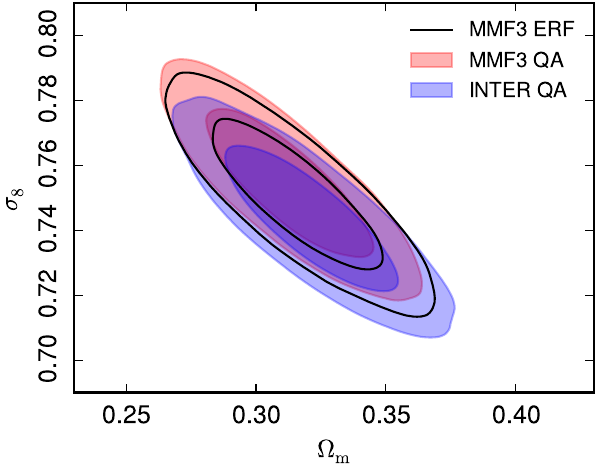}
\caption{Robustness to the choice of cluster sample and selection function model.  The shaded contours give the cosmological constraints from the 2015 MMF3 cluster catalogue using the analytical (error function, ERF) selection function model (grey), the MMF3 Monte Carlo selection function (red), and the Monte Carlo selection function for the intersection sample (blue).  Our final constraints are obtained from the MMF3 ERF model.  For this comparison, we adopt the SZ+BAO+BBN data set and we fix $1-b=0.8$.}
\label{fig:erf_qa}
\end{figure}

For our baseline analysis, we use the MMF3 cosmological sample and its associated completeness based on the analytical approximation  using the error function (Eq.~\ref{eq:completeness}).  In Fig.~\ref{fig:erf_qa} we show how the Monte Carlo determined selection function changes the cosmological constraints (labelled QA for ``Quality Assessment'' in the figure).  We also present the constraints obtained from the intersection sample defined in Sec.~\ref{sec:sampledef}.  The figure is based on the 1D $N(z)$ likelihood, for which the Monte Carlo completeness can be easily computed, and we use the baseline SZ+BAO+BBN data set and fix $1-b=0.8$. The MMF3 ERF contour is thus close to the $q=6$ contour of Fig.~\ref{fig:psz2_vs_threshold}, the only difference being that $\sigma_{\ln Y}$ is fixed to zero in Fig. A.1. while it is constrained by the Table~\ref{tab:params} prior in Fig.~\ref{fig:psz2_vs_threshold}.  The impact of adopting the intersection sample and/or the QA completeness function is small ($<0.5\,\sigma)$ for both $\OmM$ and $\sigma_8$.

\subsection{Mass function}
\label{sec:rob_mf}

\begin{figure}
\centering
\includegraphics[width=8cm]{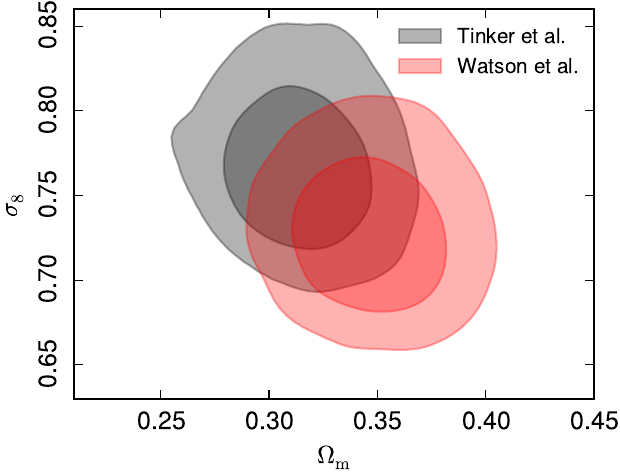}
\caption{Robustness to the choice of mass function.  The grey shaded contours give the cosmological constraints when using the~\cite{tinker2008} mass function, corresponding to our final result.  This is compared to constraints obtained when using the~\cite{watson2013} mass function, shown as the red shaded contours.  In this figure we adopt the SZ+BAO+BBN data set and the CCCP mass bias prior.}
\label{fig:mf_test}
\end{figure}

We use the~\cite{tinker2008} mass function for our baseline analysis.  To characterize the influence of this choice, we examine constraints when adopting the mass function from~\cite{watson2013} instead.  We employ our two-dimensional $N(z,q)$ likelihood (with the CCCP mass bias prior and $\alpha$ constrained) and combine with BAO and BBN prior constraints, showing the result in Fig.~\ref{fig:mf_test}.  The Tinker et al. contour of Fig.~\ref{fig:mf_test} is thus identical to the $N(z,q)$ contour with $\alpha$ constrained, as given in Fig.~\ref{fig:comp1d_2d}. The new mass function shifts our constraints by about $1 \sigma$ towards higher $\OmM$ and lower $\sigma_8$, along the main degeneracy line, hence increasing the tension with the \Planck\ primary CMB constraints.  We note that we use the general fit from Eq.~(12) of~\cite{watson2013} (independent of redshift).  This was not the case for our 2013 paper~\citep{planck2013-p15} where we adopted the AHF fit with parameters varying with redshift in the first posted version of the paper~\citep[see][on ArXiv.org]{watson2013b}, which was subsequently found to be incorrect.

\subsection{Redshift evolution of the Y-M relation}
\label{sec:rob_evol}
\begin{figure}
\centering
\includegraphics[width=8cm]{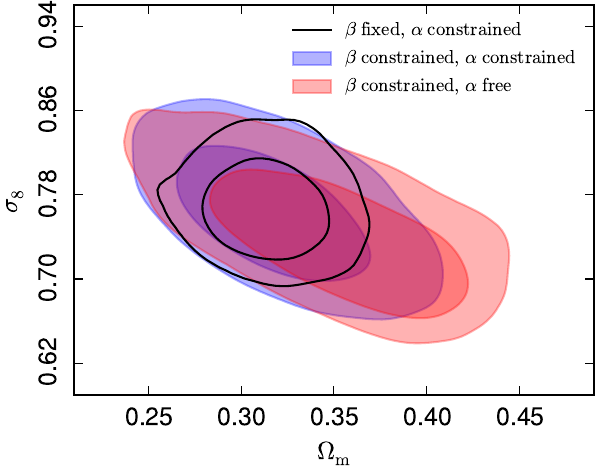}
\caption{Robustness to redshift evolution in the SZ-mass scaling relation.  The different contours show the constraints when relaxing the redshift evolution exponent, $\beta$, of Eq.~(\ref{eq:Yscaling}).  The black contours result from fixing $\beta=0.66$, our fiducial value throughout, with $\alpha$ constrained by the Gaussian X-ray prior of Table~\ref{tab:params}.  Applying a Gaussian the prior on $\beta$ instead, from Table~\ref{tab:params}, produces the blue contours, while the red contours result when we also leave $\alpha$ free.  In this figure we adopt the SZ+BAO+BBN data set and the CCCP mass bias prior. 
}
\label{fig:beta_sens}
\end{figure}

Throughout our baseline analysis, we fix the redshift evolution exponent $\beta=0.66$ (self-similar prediction) in Eq.~(\ref{eq:Yscaling}). Here we examine the impact of allowing this parameter to vary.  Constraints when leaving $\beta$ free are shown in Fig.~\ref{fig:beta_sens}. The ``$\beta$ fixed, $\alpha$ constrained'' case corresponds to the two-dimensional $N(z,q)$ likelihood (CCCP mass bias prior and $\alpha$ constrained) combined with BAO and BBN, as in Fig.~\ref{fig:mf_test}.  This contour is also identical to the $N(z,q)$ contour with $\alpha$ constrained, shown in Fig.~\ref{fig:comp1d_2d}. For the ``$\beta$ constrained'' cases, $\beta$ is allowed to vary over the range $0.66 \pm 0.50$ (Table~\ref{tab:params}).  This increases the size of our constraints along the major degeneracy between $\OmM$ and $\sigma_8$, but does not bring them into any closer agreement with the primary CMB. 

\cite{andreon2014} recently reanalyzed the subsample of 71 clusters used in~\cite{planck2013-p15}. Through a joint fit to the normalization and  redshift evolution of the Y-M relation, he found a significant detection of non-standard redshift evolution. It is possible that this conclusion is driven by systematic effects in the X-ray and SZ measurements. In particular, low-z objects have larger angular sizes, and so measurement of their X-ray and SZ quantities are subject to different systematic effects compared to equivalent measurements for high-z objects. Appendix D of~\cite{planck2013-p05a} includes a comprehensive discussion of the various systematic effects that may have an impact on measurement of the Y-M relation. Given these caveats, we believe that it is premature to draw definitive conclusions on the evolution of the Y-M relation.

\subsection{Slope of the $Y-M$ relation}
The one-dimensional analysis of the redshift distribution in Sect.~\ref{sec:12Dcomp} preferred steeper values of the $Y-M$ scaling exponent, $\alpha$, than indicated by X-ray studies of local clusters (see Fig.~\ref{fig:comp1d_2d}, case $\alpha$ free, black contours and curves).  This seems to be related to the flattening in the redshift distribution around $z=0.2$ seen in Figures~\ref{fig:psz2_dndz} and \ref{fig:psz2_dndz_comp}.  To explore this, we separately extracted parameter constraints from low ($z<0.2$) and high ($z>0.2$) redshift bin sets (two and eight bins, respectively) when leaving $\alpha$ free.

\begin{figure}
\centering
\includegraphics[width=8cm]{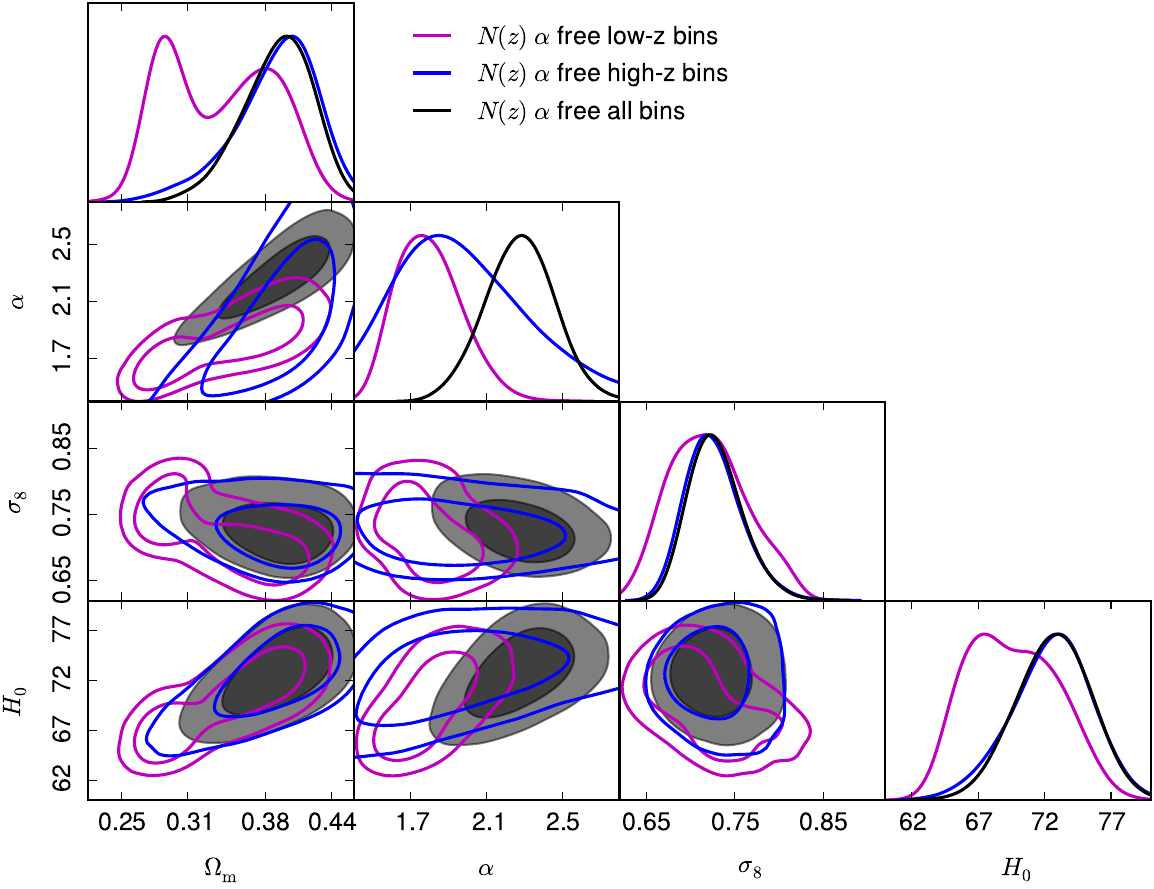} 
\caption{Comparison of constraints from the full redshift distribution (in black) to those from the redshift split: two bins at $z<0.2$ (in purple) and eight bins at $z>0.2$ (in blue). The constraints are obtained with the one-dimensional likelihood when leaving $\alpha$ free, and the black curves and contours reproduce those from Fig.~\ref{fig:comp1d_2d}.}
\label{fig:alpha}
\end{figure}

Figure~\ref{fig:alpha} compares the constraints from these splits to those from the full bin set (with $\alpha$ free).  Neither the low nor the high redshift bin set on its own prefers the steep slope -- the posteriors peak near the X-ray value, i.e, the clusters in each subsample follow the expected scaling relation.  Cosmological constraints from the low-$z$ data broaden compared to the full data set, and bimodality appears in the distribution for $\OmM$.  The high-$z$ data, on the other hand, loose precision on $\alpha$, although the peak in the posterior shifts to the X-ray value, but maintain constraints on the cosmological parameters that are similar to those from the full data set.

Only the combined data covering the full redshift range prefer the steep slope.  The parameters are coupled in a subtle interplay that favours high values of $\alpha$ and moves the cosmological parameters along the degeneracy ridges to settle on their high-$z$ values.  These degeneracies mask the interplay and unfortunately make further exploration difficult with the present data set.

\raggedright 
\end{document}